\documentclass[acmsmall,screen]{acmart}

\usepackage{amsthm, amsmath, tikz}
\usepackage{mathrsfs}
\usepackage{braket}
\usepackage{qcircuit}
\usepackage{tikz-cd}
\usepackage{hyperref}
\usepackage{stmaryrd}
\usepackage[capitalize,nameinlink,sort,noabbrev]{cleveref}
\usepackage{algorithm}
\usepackage{algpseudocode}
\usepackage{booktabs}
\usepackage{caption}
\usepackage{subcaption}
\usepackage{syntax}
\usepackage{galois}
\usepackage{nicematrix}
\usepackage{mathbbol}
\usepackage{wrapfig}
\usepackage{pdflscape}
\usetikzlibrary{positioning}

\usepackage{titlesec}
\titlespacing*{\paragraph}{0pt}{1ex plus 1ex minus .2ex}{1em}

\DeclareSymbolFontAlphabet{\mathbb}{AMSb}
\DeclareSymbolFontAlphabet{\mathbbl}{bbold}

\DeclareMathSymbol{\rel}{\mathrel}{bbold}{\lq\;}
\newcommand{\Z}{\mathbb{Z}}

\newcommand{\C}{\mathbb{C}}

\newcommand{\F}{\mathbb{F}}
\newcommand{\R}{\mathbb{R}}

\renewcommand{\H}{\mathcal{H}}

\newcommand{\cxgate}{\textsc{cnot}}
\newcommand{\ccxgate}{\textsc{tof}}
\newcommand{\swapgate}{\textsc{swap}}

\newcommand{\cczgate}{\textsc{ccz}}
\newcommand{\pgate}{\textsc{s}}
\newcommand{\zgate}{\textsc{z}}
\newcommand{\rzgate}{\textsc{r}_Z}
\newcommand{\rxgate}{\textsc{r}_X}

\newcommand{\hgate}{\textsc{h}}
\newcommand{\tgate}{\textsc{t}}
\newcommand{\xgate}{\textsc{x}}
\newcommand{\igate}{\textsc{i}}

\newcommand{\supp}[1]{\mathrm{supp}(#1)}
\newcommand{\spn}[1]{\mathrm{span}(#1)}

\newcommand{\powerset}[1]{\mathcal{P}(#1)}

\newcommand{\ketbra}[2]{|#1\rangle\langle #2|}
\newcommand{\ideal}{\mathbb{I}}
\newcommand{\variety}{\mathbb{V}}
\newcommand{\subspace}{\mathcal{S}}
\newcommand{\sep}{;\;}

\newcommand{\sem}[1]{\llbracket #1 \rrbracket}
\newcommand{\tsem}[1]{\mathcal{T}\llbracket #1 \rrbracket}

\newcommand{\abssem}[1]{\mathcal{A}\llbracket #1 \rrbracket} 
\newcommand{\qlist}{\vec q}

\newcommand{\defeq}{:=}
\newcommand{\Hilb}{\mathcal{H}}

\newcommand{\compose}{\rel}
\newcommand{\dual}[1]{(#1)^*}
\newcommand{\kleene}{\star}
\newcommand{\ps}[1]{\llparenthesis #1 \rrparenthesis}

\newtheoremstyle{break}% name
  {}%         Space above, empty = `usual value'
  {}%         Space below
  {\itshape}% Body font
  {}%         Indent amount (empty = no indent, \parindent = para indent)
  {\bfseries}% Thm head font
  {.}%        Punctuation after thm head
  {\newline}% Space after thm head: \newline = linebreak
  {}%         Thm head spec

\theoremstyle{plain}
\newtheorem{theorem}{Theorem}

\newtheorem{proposition}[theorem]{Proposition}
\newtheorem*{proposition*}{Proposition}

\theoremstyle{break}

\theoremstyle{definition}
\newtheorem{definition}[theorem]{Definition}

\newtheorem{example}[theorem]{Example}
\theoremstyle{remark}
\newtheorem{remark}[theorem]{Remark}

\title{Linear and non-linear relational analyses for Quantum Program Optimization}

\author{Matthew Amy}
\affiliation{%
  \institution{School of Computing Science, Simon Fraser University}
  \country{Canada}
}
\email{matt_amy@sfu.ca}

\author{Joseph Lunderville}
\affiliation{%
  \institution{School of Computing Science, Simon Fraser University}
  \country{Canada}
}

\date{\today}

\begin{document}

% ------------------------------------------------------------------
\begin{abstract}
The phase folding optimization is a circuit optimization used in many quantum compilers as a fast and effective way of reducing the number of high-cost gates in a quantum circuit. However, existing formulations of the optimization rely on an exact, linear algebraic representation of the circuit, restricting the optimization to being performed on straightline quantum circuits or basic blocks in a larger quantum program.

We show that the phase folding optimization can be re-cast as an \emph{affine relation analysis}, which allows the direct application of classical techniques for affine relations to extend phase folding to quantum \emph{programs} with arbitrarily complicated classical control flow including nested loops and procedure calls. Through the lens of relational analysis, we show that the optimization can be powered-up by substituting other classical relational domains, particularly ones for \emph{non-linear} relations which are useful in analyzing circuits involving classical arithmetic. To increase the precision of our analysis and infer non-linear relations from gate sets involving only linear operations ---  such as Clifford+$\tgate$ --- we show that the \emph{sum-over-paths} technique can be used to extract precise symbolic transition relations for straightline circuits. Our experiments show that our methods are able to generate and use non-trivial loop invariants for quantum program optimization, as well as achieve some optimizations of common circuits which were previously attainable only by hand.

\end{abstract}

\maketitle

\section{Introduction}\label{sec:intro}

The optimization of quantum programs is an increasingly important part of quantum computing research. On the one hand, quantum computers are scaling towards regimes of practical utility, and with it compilation tool chains are increasing in complexity. On the other hand, while the theoretical applications of quantum computers have been well established, the compilation to \emph{fault-tolerant architectures} which is necessary for such applications due to the intrinsically high error rates of quantum processors induces a massive amount of time and space overhead. In order to bring these overheads down to levels where a quantum algorithm can outperform a classical one in practical regimes, researchers have spent much effort optimizing quantum circuits and programs.

A major driver of the overhead is the high cost of \emph{magic state distillation} needed to implement certain gates. In fault-tolerant quantum error correction (FTQEC), logical gates --- gates which act on the encoded logical state rather than physical qubits --- are typically divided into those that can be implemented directly on the code space via a small number of physical gates, and those which require \emph{gate teleportation} \cite{gc99} to implement. A classic result in quantum computing \cite{ek09} states that at least one operation of the latter type is needed to achieve approximate universality. In standard codes, the \emph{Clifford group} generated by the $\hgate, \pgate,$ and $\cxgate$ gates can be implemented efficiently on the code space by \emph{transversal} operations \cite{ek09}, \emph{braiding} \cite{fmmc12}, or \emph{lattice surgery} \cite{l19} in modern schemes. The non-Clifford $\tgate\defeq \mathrm{diag}(1, \omega:= e^{\frac{i\pi}4})$ gate is typically chosen as the additional gate needed for universality, and is implemented by using magic state distillation to produce a high-fidelity $\ket{\tgate}:=\tgate\hgate\ket{0}$ state and then teleporting it into a $\tgate$ gate. As a result, the physical footprint of a $\tgate$ gate is often orders of magnitude larger than other gates \cite{l19}.

A common problem for compilers targeting fault-tolerant quantum computation is hence to reduce the number of $\tgate$ gates in a \emph{Clifford+$T$} quantum circuit, called its \emph{$\tgate$-count}, and many methods \cite{amm14,am19,nrscm17,kv20,zc19,hc18,bbw20,tensor} have been devised to do so. One standard method originating in \cite{amm14} is colloquially known as \emph{phase folding}, which optimizes circuits by finding pairs of $\tgate$ gates which can be canceled out or otherwise merged into Clifford gates via the identity $\tgate^2 = \pgate$. The optimization is crucially \emph{efficient} and \emph{monotone}, in that it operates in polynomial time in the size of the circuit, and is \emph{non-increasing on any meaningful cost metric}. As a fast, effective, and monotone optimization, many existing quantum compilers \cite{staq,voqc,tket,qiskit2024,pyzx} implement some variant of this algorithm. However, the optimization and its variants rely on the linear-algebraic semantics of a quantum circuit, and hence can only be applied to \emph{circuits}, not \emph{programs}. As quantum tool chains scale up, intermediate representations are necessarily leaving the strictly quantum-centric view and incorporating classical computation directly into compiled code \cite{qasm}, making the integration of such circuit optimizations challenging at best and ineffective at worst. 

In this work we develop a phase folding algorithm which applies to quantum \emph{programs} which involve both classical and quantum computation. We do this by re-framing the phase folding algorithm as a relational analysis which computes a sound approximation of the classical transitions in a quantum program via an affine subspace defined over the pre- and post-state. In this way we can directly apply existing techniques for affine relation analysis (ARA) such as Karr's seminal analysis \cite{karr76}. Moreover, our optimization can handle arbitrarily complicated classical control including nested loops and procedure calls, as it amounts to a process of \emph{summarization}. Framing the phase folding algorithm as an approximation of the \emph{affine} relations between the pre- and post-states in a program further allows us to generalize the optimization to  \emph{non-linear} relations. We give a non-linear relational analysis for quantum programs which abstracts the classical transitions as an affine variety over $\F_2$, represented via the reduced Gr\"{o}bner basis of a polynomial ideal. As most common quantum gates implement affine transitions in the classical state space and generate non-linear behaviour via \emph{interference}, we further develop a method of generating precise non-linear transition formulas for the basic blocks in a control-flow graph via \emph{symbolic path integrals}.

\subsection{Contributions}
In summary we make the following contributions:
\begin{itemize}
	\item We show that quantum phase folding is effectively an affine relation analysis (\cref{sec:ara}).
	\item We give the first phase folding algorithm which applies non-trivially to \emph{programs} involving both quantum and classical computation  (\cref{sec:ara}).
	\item We give a novel phase folding algorithm which makes use of \emph{non-linear} relations and is again applicable to general programs. We do so by showing that over $\F_2$, precise polynomial relations are relatively simple to compute via Gr\"{o}bner basis methods  (\cref{sec:pra}).
	\item We show that (relatively) precise transition relations can be extracted from the symbolic path integral of a basic block and used to increase precision in either domain  (\cref{sec:rewrite}).
	
	\item Conceptually, we demonstrate that \emph{classical} program analysis techniques may be productively applied to \emph{classical data in superposition}. This is in contrast to most existing work on quantum program analysis, which treat \emph{quantum data} with \emph{quantum domains}. 
\end{itemize}

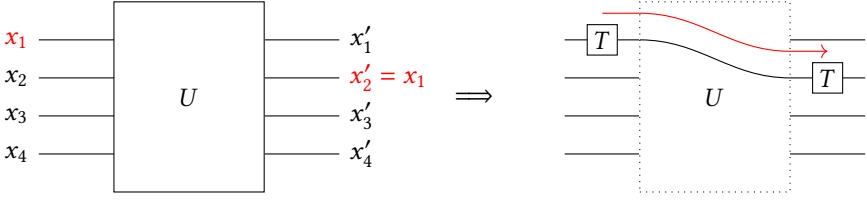
\begin{figure}
\begin{tikzpicture}
	\draw[] (1,0) -- (5,0) node{};
	\draw[] (1,0.5) -- (5,0.5) node{};
	\draw[] (1,1) -- (5,1) node{};
	\draw[] (1,1.5) -- (5,1.5) node{};
	\node[] at (0.7,1.5) {\color{red}$x_1$};
	\node[] at (0.7,1) {$x_2$};
	\node[] at (0.7,0.5) {$x_3$};
	\node[] at (0.7,0) {$x_4$};
	\node[] at (5.3,1.5) {$x_1'$};
	\node[] at (5.65,1) {\color{red}$x_2'=x_1$};
	\node[] at (5.3,0.5) {$x_3'$};
	\node[] at (5.3,0) {$x_4'$};
	\draw[fill=white] (2,-0.5) rectangle (4,2) node[pos=.5] {$U$};
	\node[] at (6.8,0.75) {$\implies$};
	\draw[] (8,0) -- (12,0) node{};
	\draw[] (8,0.5) -- (12,0.5) node{};
	\draw[] (8,1) -- (12,1) node{};
	\draw[] (8,1.5) -- (12,1.5) node{};
	\draw[fill=white, dotted] (9,-0.5) rectangle (11,2) node[pos=0.5] {$U$};
	\draw[] (9,1.5) to[out=0,in=180] (11,1) node{};
	\draw[fill=white] (8.3,1.7) rectangle (8.7,1.3) node[pos=.5] {$T$};
	\draw[fill=white] (11.3,1.2) rectangle (11.7,0.8) node[pos=.5] {$T$};
	\draw[color=red] (8.5,1.85) to[out=0,in=180] (9,1.85) node{};
	\draw[color=red] (9,1.85) to[out=0,in=180] (11,1.35) node{};
	\draw[->,color=red] (11,1.35) to[out=0,in=180] (11.5,1.35) node{};
\end{tikzpicture}
\caption{The relational approach to phase folding. If every classical state $\ket{\vec x'}$ in the support of $U\ket{\vec x}$ satisfies $x_j'=x_i$, a phase gate on $\ket{x_i}$ can be commuted through $U$. More generally, a phase conditional on some function $f:\F_2^n\rightarrow \F_2$ commutes with $U$ if and only if $f(\vec{x}) = g(\vec{x}')$ for all $\vec{x}'$ in the support of $U\ket{\vec x}$.}\label{fig:overview}
\end{figure}

\section{Overview}

\Cref{fig:overview} gives the high-level intuition of our approach. We recall\footnote{A full introduction to quantum computation and circuits can be found in \cref{sec:prelim}.} that a \emph{qubit} --- a quantum bit --- in a pure state is described as a unit vector $\ket{\psi}\in\C^2$. A \emph{quantum gate} $U$ acts on the combined state $\ket{\psi}\in \C^{2^n}$ of $n$ qubits as a unitary (i.e. invertible) linear transformation $\ket{\psi} \mapsto U\ket{\psi}$. A \emph{quantum circuit} is a sequence of quantum gates acting on subsets of the qubits, drawn as a \emph{circuit diagram} where horizontal wires carry qubits from left to right into gates. We fix a basis $\{\ket{\vec x} \mid \vec x \in \F_2^n:=\{0,1\}^n\}$ of $\C^{2^n}$ and call these the \emph{classical states}; a state $\ket{\psi} = \sum_{\vec x \in \F_2^n} a_{\vec x}\ket{\vec x}$ is hence in a \emph{superposition} of the classical states $\{\vec x \mid a_{\vec x} \neq 0\}$, called the \emph{classical support} of $\ket{\psi}$. We can then describe the \emph{classical} semantics of a quantum gate or circuit as the \emph{classical} states which are in superposition after applying the transformation. \emph{Our methods are based around approximating the classical semantics of a quantum circuit as a set of relations or constraints which hold between $\vec x$ and the support of $U\ket{\vec x}$}. If certain relations hold, then \emph{phase gates} --- gates which are diagonal in the classical basis --- can be commuted through the program as in \cref{fig:overview} and canceled or merged with other phase gates --- a type of \emph{quantum constant propagation}.

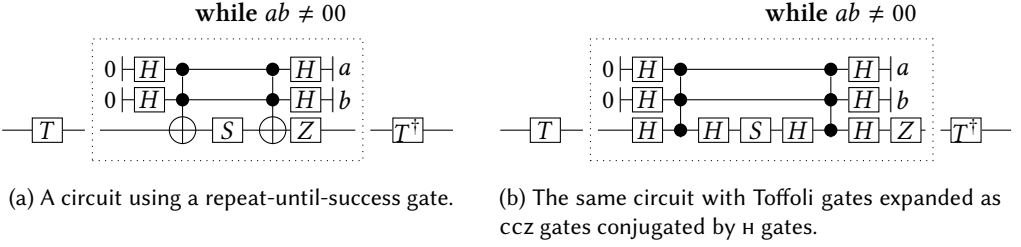
\begin{figure}
\centering
\begin{subfigure}[t]{0.45\textwidth}
\centering
\begin{tikzpicture}[scale=0.8]
	\draw[] (1,0) -- (8.5,0) node{};
	\draw[fill=white] (1.5,-0.2) rectangle (2,0.2) node[pos=.5] {$T$};
	\draw[fill=white] (7.5,-0.2) rectangle (8,0.2) node[pos=.5] {$T^\dagger$};
	\node[] at (2.8,0.5) {$0$};
	\node[] at (2.8,1) {$0$};
	\draw[] (3,0.5) node{|} -- (6.5,0.5) node{|};
	\draw[] (3,1) node {|} -- (6.5,1) node{|};
	\node[] at (6.7,0.5) {$b$};
	\node[] at (6.7,1) {$a$};
	\draw[fill=white] (3.2,0.8) rectangle (3.7,1.2) node[pos=.5] {$H$};
	\draw[fill=white] (3.2,0.3) rectangle (3.7,0.7) node[pos=.5] {$H$};
	\node[circle,fill=black,inner sep=0pt,minimum size=5pt] at (4,0.5) {};
	\node[circle,fill=black,inner sep=0pt,minimum size=5pt] (a) at (4,1) {};
	\node[circle,draw=black,inner sep=0pt,minimum size=10pt] (b) at (4,0) {};
	\draw[fill=white] (4.5,-0.2) rectangle (5,0.2) node[pos=.5] {$S$};
	\draw[] (a) -- (b.south) node {};
	\node[circle,fill=black,inner sep=0pt,minimum size=5pt] at (5.5,0.5) {};
	\node[circle,fill=black,inner sep=0pt,minimum size=5pt] (c) at (5.5,1) {};
	\node[circle,draw=black,inner sep=0pt,minimum size=10pt] (d) at (5.5,0) {};
	\draw[] (c) -- (d.south) node {};
	\draw[fill=white] (5.8,0.8) rectangle (6.3,1.2) node[pos=.5] {$H$};
	\draw[fill=white] (5.8,0.3) rectangle (6.3,0.7) node[pos=.5] {$H$};
	\draw[fill=white] (5.8,-0.2) rectangle (6.3,0.2) node[pos=.5] {$Z$};
	\draw[draw=white, line width = 2mm] (2.5,-0.5) rectangle (7,1.5) node[above left=0mm] {$\mathbf{while}\;  ab\neq 00$};
	\draw[dotted] (2.5,-0.5) rectangle (7,1.5) node[pos=.5] {};
\end{tikzpicture}
\caption{A circuit using a repeat-until-success gate.}\label{fig:rus}
\end{subfigure}
\quad
\begin{subfigure}[t]{0.48\textwidth}
\centering
\begin{tikzpicture}[scale=0.8]
	\draw[] (1,0) -- (9.5,0) node{};
	\draw[fill=white] (1.5,-0.2) rectangle (2,0.2) node[pos=.5] {$T$};
	\draw[fill=white] (8.5,-0.2) rectangle (9,0.2) node[pos=.5] {$T^\dagger$};
	\node[] at (2.8,0.5) {$0$};
	\node[] at (2.8,1) {$0$};
	\draw[] (3,0.5) node{|} -- (7.5,0.5) node{|};
	\draw[] (3,1) node {|} -- (7.5,1) node{|};
	\node[] at (7.7,0.5) {$b$};
	\node[] at (7.7,1) {$a$};
	\draw[fill=white] (3.2,0.8) rectangle (3.7,1.2) node[pos=.5] {$H$};
	\draw[fill=white] (3.2,0.3) rectangle (3.7,0.7) node[pos=.5] {$H$};
	\draw[fill=white] (3.2,-0.2) rectangle (3.7,0.2) node[pos=.5] {$H$};
	\node[circle,fill=black,inner sep=0pt,minimum size=5pt] at (4,0.5) {};
	\node[circle,fill=black,inner sep=0pt,minimum size=5pt] (a) at (4,1) {};
	\node[circle,fill=black,inner sep=0pt,minimum size=5pt] (b) at (4,0) {};
	\draw[fill=white] (4.3,-0.2) rectangle (4.8,0.2) node[pos=.5] {$H$};
	\draw[fill=white] (5,-0.2) rectangle (5.5,0.2) node[pos=.5] {$S$};
	\draw[fill=white] (5.7,-0.2) rectangle (6.2,0.2) node[pos=.5] {$H$};
	\draw[] (a) -- (b.south) node {};
	\node[circle,fill=black,inner sep=0pt,minimum size=5pt] at (6.5,0.5) {};
	\node[circle,fill=black,inner sep=0pt,minimum size=5pt] (c) at (6.5,1) {};
	\node[circle,fill=black,inner sep=0pt,minimum size=5pt] (d) at (6.5,0) {};
	\draw[] (c) -- (d.south) node {};
	\draw[fill=white] (6.8,0.8) rectangle (7.3,1.2) node[pos=.5] {$H$};
	\draw[fill=white] (6.8,0.3) rectangle (7.3,0.7) node[pos=.5] {$H$};
	\draw[fill=white] (6.8,-0.2) rectangle (7.3,0.2) node[pos=.5] {$H$};
	\draw[fill=white] (7.5,-0.2) rectangle (8,0.2) node[pos=.5] {$Z$};
	\draw[draw=white, line width = 2mm] (2.5,-0.5) rectangle (8.2,1.5) node[above left=0mm] {$\mathbf{while}\;  ab\neq 00$};
	\draw[dotted] (2.5,-0.5) rectangle (8.2,1.5) node[pos=.5] {};
\end{tikzpicture}
\caption{The same circuit with Toffoli gates expanded as $\cczgate$ gates conjugated by $\hgate$ gates.}\label{fig:rus2}
\end{subfigure}
\caption{Repeat-Until-Success circuits. The box denotes a loop which  terminates exactly when the measurement result $xy$ of the first two qubits is $00$.}\label{fig:ruswhole}
\end{figure}

To illustrate our methods, consider the circuit in \cref{fig:rus} which uses a \emph{repeat-until-success} (RUS) implementation of the gate $\textsc{v}=(\igate + 2i\zgate)/\sqrt{5} = \mathrm{diag}(\frac{1 + 2i}{\sqrt{5}}, \frac{1-2i}{\sqrt{5}})$ from \cite{nc00}. The RUS circuit, corresponding to the boxed loop, implements $\textsc{v}$ by repeatedly initializing two additional (\emph{ancilla}) qubits in the $\ket{0}$ state, performing a computation, then measuring the ancilla qubits to get $a,b\in\F_2$. The resulting transformation on the target qubit after every iteration of the loop is a diagonal gate: $\igate$ if $ab\neq 00$, and $\textsc{v}$ if $ab=00$. The (diagonal) $\tgate$ gate before entering the loop can thus cancel out with the $\tgate^\dagger$ at the end of the loop, as for any diagonal operators $U$,$V$ we have $UV=VU$. 

Achieving this optimization is challenging for existing methods, as the circuit has no finite interpretation as a linear operator. Instead, our method works by generating a (classical) \emph{loop invariant}, namely that for any initial state $\ket{z}$ where $z\in\F_2$, the state at termination has support $\{z\}$ --- i.e. the loop is diagonal in the computational basis. 
To achieve this, we view the classical semantics of a gate as a non-deterministic transition relation $R\subseteq \F_2^n\times \F_2^n$ on the classical states and compute a transition for the entire loop body by composing each relation. The relations for the gates involved are given below, where $\ket{x}\mapsto \ket{x'}$ denotes the relation $\{(x,x') \mid x,x'\in \F_2\}$: 
\[
	\raisebox{0.13em}{\footnotesize $\Qcircuit @C=.4em @R=.001em @! { & \gate{H} & \qw }$}=\hgate:\ket{x} \mapsto \ket{x'} \qquad\qquad \raisebox{0.7em}{\footnotesize $\Qcircuit @C=.4em @R=.001em @! { & \ctrl{2} & \qw \\ &  \ctrl{1} & \qw \\ &  \targ & \qw}$}=\ccxgate:\ket{x,y,z}\mapsto \ket{x,y,z\oplus (x\land y)} 
\] 
\[
	\raisebox{0.13em}{\footnotesize $\Qcircuit @C=.4em @R=.001em @! { & \gate{S} & \qw }$}=\pgate:\ket{x} \mapsto \ket{x}\qquad\qquad \raisebox{0.13em}{\footnotesize $\Qcircuit @C=.4em @R=.001em @! { & \gate{Z} & \qw }$}=\zgate:\ket{x} \mapsto \ket{x}.
\] 
Composing the relations corresponding to each gate on an initial state $\ket{z}$ and we have
\begin{equation*}
\resizebox{0.95\textwidth}{!}{$
	\ket{z} \xrightarrow{\mathrm{init}} \ket{0,0,z} \xrightarrow{\hgate\otimes \hgate\otimes \igate} \ket{x,y,z} \xrightarrow{\ccxgate} \ket{x,y,z\oplus (x\land y)} \xrightarrow{\igate\otimes\igate\otimes \pgate} \ket{x,y,z\oplus (x\land y)} \xrightarrow{\ccxgate} \ket{x,y,z} \xrightarrow{\hgate\otimes \hgate\otimes \zgate} \ket{x',y',z} \xrightarrow{\mathrm{meas}} \ket{z}$
}
\end{equation*}
We then take \emph{Kleene closure} of the loop body transition $\mathrm{RUS}_{\text{body}}:\ket{z} \mapsto \ket{z}$, which approximates (in this case, precisely) the composition $\mathrm{RUS}_{\text{body}}^k$ of \emph{any number $k$ of loop iterations} as $\ket{z} \mapsto \ket{z}$. Hence for any number of iterations of the loop $\mathrm{RUS}_{\text{body}}$, the state of the qubit in the computational basis is unchanged and so the $\tgate$ gate commutes through and cancels out the $\tgate^\dagger = \tgate^{-1}$ gate.

The problem becomes more challenging when the circuit is written in the Clifford+$\tgate$ gate set, or similar gate sets such as Clifford+$\cczgate$ which implement classical non-linearity via \emph{interference}. \Cref{fig:rus2} gives an alternative implementation of the RUS circuit over the Clifford+$\cczgate$. The $\ccxgate$ gates have been replaced with $\raisebox{0.5em}{\footnotesize $\Qcircuit @C=.4em @R=.2em @! { & \ctrl{2} & \qw \\ &  \ctrl{1} & \qw \\ &  \ctrl{0} & \qw}$}=\cczgate$ gates conjugated by Hadamard gates, via the circuit equality $\ccxgate = (\igate\otimes \igate\otimes \hgate)\cczgate(\igate\otimes \igate\otimes \hgate)$. The $\cczgate$ is diagonal, and hence its classical semantics is given by the identity relation $\cczgate:\ket{x,y,z} = \ket{x,y,z}$. Composing again the relations for each gate, we now find that the loop body has the following classical relation
\[
	\mathrm{RUS}_{\text{body}}:\ket{z} \mapsto \ket{z'}.
\]
The relation above over-approximates the precise loop invariant $\ket{z}\mapsto \ket{z}$. Over-approximation is due to the fact that the relation $(\igate\otimes \igate\otimes \hgate)\cczgate(\igate\otimes \igate\otimes \hgate):\ket{x,y,z}\mapsto \ket{x,y,z'}$ over-approximates the precise semantics of the $\ccxgate$ gate, namely that $z' = z\oplus (x\land y)=z \oplus xy$, itself a consequence of the quantum mechanical effect of \emph{interference} which can cause classical transitions to \emph{cancel}.

To recover the precise classical semantics, we use \emph{path integrals} to analyze interference and compute accurate transition relations. Computing the path integral for $(\igate\otimes \igate\otimes \hgate)\cczgate(\igate\otimes \igate\otimes \hgate)$ gives the \emph{quantum} transition relation:
\[
	(\igate\otimes \igate\otimes \hgate)\cczgate(\igate\otimes \igate\otimes \hgate):\ket{x,y,z} \mapsto \frac{1}{2}\sum_{z'}\left(\sum_{z''\in\F_2} (-1)^{zz''\oplus  xyz'' \oplus z''z'}\right)\ket{x,y,z'}.
\]
From the expression above we see that whenever $z'\neq z \oplus xy$ in the outer sum, the \emph{paths} corresponding to the inner sum over $z''$ have opposite phase $(\pm 1)$ and hence cancel out, so the circuit satisfies the additional relation $z \oplus xy \oplus z' = 0$. Performing this analysis for the loop in \cref{fig:rus2} we get the following system of equations, where $t_1,\dots, t_4$ denote intermediate variables corresponding to the outputs of the $4$ $\hgate$ gates on the target qubit:
\begin{align*}
	z \oplus xy \oplus t_2 = 0 \qquad\quad
	t_2 \oplus xy \oplus t_4 = 0 \quad\qquad
	z' \oplus t_4 = 0
\end{align*}
Solving the system of polynomial equations above for $z'$ over the input variable $z$ we find that $z' = z$, as was the case for \cref{fig:rus}. While simple back substitution suffices in this case, in more general cases the primed output variables may not have a simple solution in terms of the input variables (or indeed, may have no solution at all). We solve the system of polynomial equations by computing a \emph{Gr\"{o}bner basis} of the ideal generated by the system of equations, and then using \emph{elimination theory} to project out temporary variables. Doing so for the above system of equations produces the polynomial ideal $\langle z \oplus z' \rangle\subseteq \F_2[z,z']$ which in particular implies that $z' = z$.

\begin{figure}
\begin{tikzpicture}
	\draw[] (0,0.5) -- (3,0.5) node{};
	\draw[] (0,0) -- (3,0) node{};
	\node[circle,fill=black,inner sep=0pt,minimum size=5pt] (a) at (0.5,0.5) {};
	\node[circle,draw=black,inner sep=0pt,minimum size=10pt] (b) at (0.5,0) {};
	\draw[] (a) -- (b.south) node {};
	\draw[fill=white] (1,-0.2) rectangle (1.5,0.2) node[pos=.5] {$T$};
	\node[circle,fill=black,inner sep=0pt,minimum size=5pt] (c) at (2,0.5) {};
	\node[circle,draw=black,inner sep=0pt,minimum size=10pt] (d) at (2,0) {};
	\draw[] (c) -- (d.south) node {};
	\draw[] (3,0.5) to[out=0,in=180] (4,0) node{};
	\draw[] (3,0) to[out=0,in=180] (4,0.5) node{};
	\draw[] (4,0.5) -- (7,0.5) node{};
	\draw[] (4,0) -- (7,0) node{};
	\node[circle,fill=black,inner sep=0pt,minimum size=5pt] (e) at (5,0.5) {};
	\node[circle,draw=black,inner sep=0pt,minimum size=10pt] (f) at (5,0) {};
	\draw[] (e) -- (f.south) node {};
	\draw[fill=white] (5.5,-0.2) rectangle (6,0.2) node[pos=.5] {$T^\dagger$};
	\node[circle,fill=black,inner sep=0pt,minimum size=5pt] (g) at (6.5,0.5) {};
	\node[circle,draw=black,inner sep=0pt,minimum size=10pt] (h) at (6.5,0) {};
	\draw[] (g) -- (h.south) node {};
	\draw[draw=white, line width = 2mm] (2.5,-0.5) rectangle (4.5,1) node[above left=0mm] {$\mathbf{while}\;  \star$};
	\draw[dotted] (2.5,-0.5) rectangle (4.5,1) node[pos=.5] {};
\end{tikzpicture}
\caption{A circuit with eliminable $\tgate$ gates. The strongest affine loop invariant on the classical support is the relation $x'\oplus y' = x\oplus y$, which implies that the total phase contribution of both $\tgate$ gates is $1$.}\label{fig:cool}
\end{figure}
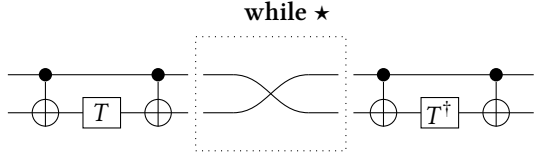

In the case of the RUS circuit, the loop invariant is simply $z' = z$, but in more exotic cases we can make use of more interesting loop invariants to eliminate of merge phase gates. Consider the circuit in \cref{fig:cool} which contains a single non-deterministic while loop which swaps the two qubits. It can be observed that as a function of the classical basis, the loop body either sends $\ket{x,y} \mapsto \ket{x,y}$ or to $\ket{y,x}$ depending on the parity of the number of iterations, and so neither $x'$ nor $y'$ has a solution over the input state. However, it can be observed that the relation $x' \oplus y' = x \oplus y$ holds in either case, and is hence a classical invariant of the loop. Moreover, as this is an \emph{affine} relation we can compute the Kleene closure over a weaker (but computationally more efficient) domain of \emph{affine relations}. This invariant then suffices to eliminate both $\tgate$ gates, as they contribute conjugate phases of $\omega$ and $\overline{\omega}$ when $x\oplus y = 1$ and $x'\oplus y'=1$, respectively, hence the total phase is $1$.

\section{Preliminaries}\label{sec:prelim}

We begin by reviewing the basics of quantum computing and Dirac notation, followed by our program model and the phase folding optimization. For a more complete introduction to quantum computation, the reader is directed to \cite{nc00}.

\subsection{Quantum computing}

Let $\H$ denote a finite-dimensional Hilbert space with dimension $\dim(\H)$. We typically assume $\H=\C^{d}$ for some fixed dimension $d$. A \emph{pure} state of a quantum system is a unit vector $\ket{v}\in\H$. The Hermitian adjoint of $\ket{v}$ in $\H^*$, denoted by $\bra{v}$, is the linear functional $\bra{v}:\ket{u}\mapsto\braket{v|u}$ where $\braket{v|u}$ denotes the (Hermitian) inner product of $\ket{v}$ and $\ket{u}$. Concretely, the Hermitian adjoint is given by the conjugate transpose, $\bra{v} = \ket{v}^\dagger$ and $\braket{v|u}= \bra{v}\cdot \ket{u}$. The outer product of $\ket{v}$ and $\ket{u}$ is similarly written $\ketbra{v}{u}$. Given two vectors $\ket{v}\in\H_1$, $\ket{u}\in\H_2$ in Hilbert spaces $\H_1$, $\H_2$, we write $\ket{v}\otimes\ket{u}\in\H_1\otimes\H_2$ for their combined state, where $\otimes$ is the tensor product. If $\ket{\psi}\in \H_1\otimes\H_2$ can not be written as a tensor product of states of $\H_1$ and $\H_2$ then $\ket{\psi}$ is said to be \emph{entangled} or \emph{non-separable}. Recall that $\dim(\H_1\otimes\H_2) = \dim(\H_1)\dim(\H_2)$, and given orthonormal bases $\{\ket{e_i}\}$, $\{\ket{f_j}\}$ of $\H_1$ and $\H_2$, respectively, $\{\ket{e_i}\otimes\ket{f_j}\}$ forms an orthonormal basis of $\H_1\otimes\H_2$. We often write $\ket{v, u}$ for the tensor product $\ket{v}\otimes \ket{u}$ when the meaning is clear. Given a set of quantum variables $\mathcal{Q}$, corresponding to distinct subsystems $q$ with local Hilbert space $\Hilb_{q}$, we denote the Hilbert space of the combined system as $\Hilb_{\mathcal{Q}} \defeq \bigotimes_{q\in\mathcal{Q}} \Hilb_{q}$. Tensor factors may be rearranged at will, and so we will sometimes use the notation $\ket{\psi}_{\qlist}\otimes \ket{\varphi}_{\mathcal{Q} - \qlist}$ to denote a separable state where $\ket{\psi}$ is the state of the qubits listed in $\qlist = q_1\cdots q_k$ and $\ket{\varphi}$ is the state of the remaining qubits.

The Hilbert space associated to a 2-dimensional quantum state --- called a \emph{qubit} --- is $\C^2$ and we denote the standard or \emph{computational} basis of $\C^2$ by $\{\ket{0},\ket{1}\}$. The $n$-fold tensor product of $\C^2$ is $\C^2\otimes \C^2\otimes \cdots \otimes\C^2\simeq \C^{2^n}$ with computational basis $\{\ket{\vec{x}} \mid \vec{x}\in\F_2^n\}$ where $\ket{x_1x_2} = \ket{x_1}\otimes \ket{x_2}$ and $\F_2=(\{0,1\}, \oplus, \cdot)$ is the $2$-element finite field. Written over the computational basis, a pure state of $n$ qubits $\ket{v} = \sum_{\vec{x}\in\F_{2}^n} \alpha_{\vec{x}}\ket{\vec{x}}$ is said to be in a \emph{superposition} of the classical states $\vec{x}\in\F_{2}^n$ for which $\alpha_{\vec{x}} \neq 0$. We call this set the \emph{classical support} or just support of a pure state $\ket{v}$, denoted $\supp{\ket{v}}$. Given the pure state $\ket{v}$ above, \emph{measuring} $\ket{v}$ in the computational basis projects the state down to some $\ket{\vec{y}}\in\supp{\ket{v}}$ with probability $|\alpha_{\vec y}|^2$. More generally we may measure just a single qubit or subset of qubits --- known as a \emph{partial measurement} --- which sends a pure state $\sum_{\vec y}\alpha_{\vec y}\ket{\vec{y}}_{\qlist}\otimes \ket{\psi_{\vec y}}_{\mathcal{Q} - \qlist}$ to the pure state $\ket{\vec y}_{\qlist}\otimes \ket{\psi_{\vec y}}_{\mathcal{Q} - \qlist}$ with probability $|\alpha_{\vec y}|^2$.

\subsection{Quantum circuits}

A \emph{quantum gate} is a unitary operator $U$ on a Hilbert space $\H$. By unitary we mean that its inverse is equal to its Hermitian adjoint $U^\dagger$, i.e. $UU^\dagger=U^\dagger U = I$. A quantum gate $U$ sends a pure state $\ket{v}$ to the pure state $U\ket{v}$. If $U:\H_1\rightarrow \H_1$ and $V:\H_2\rightarrow \H_2$ are unitary gates, then their tensor product $U\otimes V:\H_1\otimes\H_2\rightarrow\H_1\otimes\H_2$ is likewise unitary. A \emph{quantum circuit} is a well-formed term over the signature $(\mathcal{G}, \cdot, \otimes, \dagger)$ where $\mathcal{G}$ is a finite set of gates. We use $U_{\qlist}$ to denote a unitary on $\H_\mathcal{Q}$ where $U$ is applied to the sub-systems $\qlist$ and the identity gate $\igate$ is applied to the remaining sub-systems. Circuits are drawn graphically as in \cref{fig:circuit}, with time flowing left to right and individual wires corresponding to qubits. Vertical composition corresponds to the tensor product $\otimes$, while horizontal composition corresponds to functional composition $(\cdot)$.
\begin{figure}
\[\Qcircuit @C=1em @R=0.7em @!R {
 & \ctrl{2} & \qw  \\
 & \ctrl{1} & \qw  \\
 & \targ & \qw 
}
\quad
\raisebox{-1.8em}{$=$}
\quad
\Qcircuit @C=1em @R=.2em {
 & \gate{T} & \targ & \gate{T^\dagger} & \targ & \gate{T} & \targ & \gate{T^\dagger} & \targ & \qw  \\
 & \gate{T} & \qw & \ctrl{1} & \ctrl{-1} & \ctrl{1} & \qw & \qw & \ctrl{-1} & \qw  \\
 & \gate{H} & \ctrl{-2} & \targ & \gate{T^\dagger} & \targ & \ctrl{-2} & \gate{T} & \gate{H} & \qw 
}
\]
\centerline{
\begin{tikzpicture}
\draw[->, thick, red] (0,0) to (9,0);
\node[red] at (4.3, -.3) {Time};
\end{tikzpicture}
}
\caption{An example of a quantum circuit, implementing the Toffoli gate $\ccxgate:\ket{x_1,x_2,x_3} \mapsto \ket{x_1,x_2,x_3\oplus x_1x_2}$}
\label{fig:circuit}
\end{figure}

\begin{figure}
\[
  \xgate \; =\; \Qcircuit @C=1em @R=.5em {&\gate{X}&\qw} \; = \; 
	\begin{bmatrix} 0&1 \\ 1&0 \end{bmatrix} \quad
  \zgate\; =\; 
	\Qcircuit @C=1em @R=.5em {&\gate{Y}&\qw} \; = \; 
	\begin{bmatrix} 1 & 0 \\ 0 & -1 \end{bmatrix} \quad
  \rzgate(\theta)\; =\; 
	\Qcircuit @C=1em @R=.5em {&\gate{R_Z(\theta)}&\qw} \; = \; 
	\begin{bmatrix} 1 & 0 \\ 0 & e^{i\theta} \end{bmatrix}
\]
\[
  \hgate\; =\; \Qcircuit @C=1em @R=.5em {&\gate{H}&\qw} \; = \; 
	\frac{1}{\sqrt{2}} \begin{bmatrix} 1 & 1 \\ 1 & -1 \end{bmatrix} \quad
  \pgate\; =\; 
	\Qcircuit @C=1em @R=.5em {&\gate{S}&\qw} \; = \; 
	\begin{bmatrix} 1 & 0 \\ 0 & i \end{bmatrix} \quad
  \tgate \; =\; 
	\Qcircuit @C=1em @R=.5em {&\gate{T}&\qw} \; = \; 
	\begin{bmatrix} 1&0 \\ 0&\omega:=e^{i\frac{\pi}{4}} \end{bmatrix}
\]
\[
  {\cxgate \;=\; \raisebox{.8em}{\Qcircuit @C=1em @R=.5em { & \ctrl{1} & \qw \\ & \targ & \qw }} \;=\; 
  \begin{bmatrix} 
    1 & 0 & 0 & 0 \\ 
    0 & 1 & 0 & 0 \\
    0 & 0 & 0 & 1 \\
    0 & 0 & 1 & 0
  \end{bmatrix}} \qquad
  {\swapgate \;=\; \raisebox{-.6em}{\begin{tikzpicture}
	\draw[] (1.8,0) to (2,0) node{};
	\draw[] (1.8,0.4) to (2,0.4) node{};
	\draw[] (2,0) to[out=0,in=180] (2.5,0.4) node{};
	\draw[] (2,0.4) to[out=0,in=180] (2.5,0) node{};
	\draw[] (2.5,0) to (2.7,0) node{};
	\draw[] (2.5,0.4) to (2.7,0.4) node{};
	\end{tikzpicture}
	} \;=\; 
  \begin{bmatrix} 
    1 & 0 & 0 & 0 \\ 
    0 & 0 & 1 & 0 \\
    0 & 1 & 0 & 0 \\
    0 & 0 & 0 & 1
  \end{bmatrix}}
\]
\caption{Standard gates and their circuit notation.}\label{fig:gates}
\end{figure}

Standard gates are shown in \cref{fig:gates}. The \emph{Clifford+$T$} gate set is defined as $\{\hgate, \cxgate, \pgate, \tgate\}$ where $\pgate = \tgate^2$ is included as an explicit generator since, being Clifford, it is typically orders of magnitude less expensive than the $\tgate$ gate. It is often illuminating to study quantum gates by their actions on \emph{classical} states, which suffices due to the linearity of quantum mechanics. The gates of \cref{fig:gates} expressed as functions of classical states are shown below. In this view, it is clear that $\xgate$ and $\cxgate$ both perform \emph{classical} functions, while the $\zgate$, $\pgate$ and $\tgate$ operate in the (orthogonal) phase space, and the $\hgate$ is a form of \emph{quantum branching} gate, sending a classical state $\ket{x}$ to both $\ket{0}$ and $\ket{1}$ with equal probability but varying phase. By orthogonal here we mean that computations in the state space (or $Z$-basis in quantum mechanical terms) don't affect the phase, and vice versa with computations in the phase space or $X$-basis. 
\begin{align*}
	\xgate:\ket{x}\mapsto \ket{1\oplus x} \qquad
	\zgate:\ket{x}\mapsto (-1)^x\ket{x} \qquad
	\rzgate(\theta):\ket{x}\mapsto e^{i\theta x}\ket{x} \\
	\hgate:\ket{x}\mapsto\frac{1}{\sqrt{2}}\sum_{y\in\F_2}(-1)^{xy}\ket{y} \qquad
	\pgate:\ket{x}\mapsto i^x\ket{x} \qquad
	\tgate:\ket{x} \mapsto \omega^x\ket{x} \\
	\cxgate:\ket{x,y}\mapsto \ket{x,x\oplus y} \qquad
	\swapgate:\ket{x,y}\mapsto \ket{y,x} \qquad
\end{align*}

The interpretation of quantum mechanical processes as \emph{classical} processes happening in superposition is central to Feynman's \emph{path integral} formulation of quantum mechanics \cite{fh65}, and the above expressions can be thought of as \emph{symbolic path integrals}.

\subsection{Programming model}

We adopt the standard QRAM model of quantum computation, where \emph{data} can be quantum or classical, but \emph{control} is strictly classical. We illustrate our techniques on a non-deterministic version of the imperative quantum WHILE language \cite{qwhile,m12} with procedures, the syntax of which is presented below:
\begin{align*}
	T\in \mathbf{QWhile}\; ::&= \mathbf{skip}  
		\; \mid \; q := \ket{0}
		\; \mid \; U\qlist
		\; \mid \; \mathbf{meas} \; q 
		\; \mid \; \mathbf{call} \; p(\qlist)
		\; \mid \; T_1; \; T_2  \\
		&\; \mid \; \mathbf{if} \; \star\; \mathbf{then}\; T_1\; \mathbf{else} \; T_2 
		\; \mid \; \mathbf{while} \; \star\; \mathbf{do} \; T\;
\end{align*}
We assume a fixed set of quantum variables $\mathcal{Q}$ and a set of primitive gates $\mathcal{G}$. Variables $q\in\mathcal{Q}$ denote qubit identifiers, while $U\in\mathcal{G}$ denote primitive gates. Note that the syntax corresponds precisely to the syntax of regular expressions over a grammar consisting of reset instructions, unitary applications, measurements in the computational basis, and procedure calls.

We are interested in \emph{relational} properties of quantum WHILE programs --- that is, properties which relate the post-state of a program $T$ to the pre-state. Such analyses often operate by computing a \emph{summary} or \emph{abstract transformer} $\sem{T}^\sharp:\mathcal{D}\rightarrow \mathcal{D}$ for the program over an abstract domain $\mathcal{D}$ of relevant properties. Tarjan \cite{tarjan81} introduced a compositional approach to these types of analyses, sometimes called \emph{algebraic program analysis}, in which summarization proceeds by first solving the \emph{path expression problem} --- representing a program's control-flow graph as a regular expression --- and then reinterpreting the regular expression over a suitable algebraic domain of abstract transformers, notably one equipped with the regular algebra notions of \emph{choice} (denoted $+$ or $\sqcup$), \emph{composition} ($\cdot$ or $\compose$), and \emph{iteration} ($\kleene$). Our methods are based on this algebraic style of program analysis, and in particular can be applied more to any control-flow graph with quantum instructions by first mapping it to a path expression via Tarjan's algorithm \cite{tarjan81b}.

To this end, we define the semantics of a quantum WHILE program in terms of control-flow paths. We define a set of \emph{actions} $\Sigma$ as either the application of a unitary or the post-selection of a variable $q$ on a classical value $x$, with a special n-op action for convenience:
\[
	\Sigma ::= \mathbf{skip}\; \mid \; U\qlist \; \mid \; \mathbf{assume} \; P_q^x
\]
A \emph{control-flow path} $\pi\in\Sigma^*$ is a sequence of actions. We denote the sequential composition of paths by $\pi \compose \pi'$ and extend $\compose$ to sets of paths. We define the semantics $\sem{\pi}:\mathcal{H}_{\mathcal{Q}} \rightarrow \mathcal{H}_{\mathcal{Q}}$  of a control-flow path $\pi$ as a composition of linear operators corresponding to the individual actions, where for instance $\sem{U\qlist} = U_{\qlist}$ and $P^x=\ketbra{x}{x}$ is the projection onto the basis state $\ket{x}$. We say a path $\pi$ is \emph{feasible} if it is non-zero as a linear operator --- that is, there exists $\ket{v}\in\Hilb_{\mathcal{Q}}$ such that $\sem{\pi}\ket{v} \neq 0$. An example of an infeasible path is one which first projects on to the $\ket{0}$ state of a qubit, then the $\ket{1}$ state: \[\sem{\mathbf{assume} \; P_q^0;\;\mathbf{assume} \; P_q^1} = P_q^1P_q^0 = 0.\]
Note that with this interpretation, a path might not send a unit vector to another unit vector (hence, pure state). As we are only interested in the \emph{support} of a path, this definition suffices for our purposes.

\Cref{fig:csemantics} defines a collecting semantics for the quantum WHILE language in terms of paths. Paths correspond to control-flow paths over two different types of classical non-determinism: classical branching corresponding to if and while statements, and branching due to measurement outcomes of quantum states. Measurements are modeled as a non-deterministic choice between the projector $P^0=\ketbra{0}{0}$ or the projector $P^1=\ketbra{1}{1}$. Reset statements are modeled as a measurement followed by an $X$ correction in the event of measurement outcome $1$. The path semantics induces a collecting semantics $\sem{T}:\powerset{\Hilb_\mathcal{Q}}\rightarrow\powerset{\Hilb_\mathcal{Q}}$ in the obvious way. It can be observed that up to normalization, $\ket{u}\in \sem{T}\{\ket{v}\}$ if and only if it results from some sequence of measurements and (non-deterministic) classical branches with non-zero probability. In particular, it can be observed that measuring a qubit which has been reset to $\ket{0}$ has no effect, as shown below where $S\in\powerset{\Hilb_{\mathcal{Q}}}$:
\[
	\sem{q := \ket{0};\;\mathbf{meas} \; q }S = \sem{q := \ket{0}}S = \{\ket{0}_q\otimes \ket{\psi}_{\mathcal{Q}-\{q\}} \mid\ket{x}_q\otimes \ket{\psi}_{\mathcal{Q}-\{q\}} \in S \}.
\]

\begin{example}
	Consider the program $Hq; \; q := \ket{0}$ which applies a Hadamard gate to qubit $q$ then resets $q$ to $\ket{0}$. Up to normalization we have
	\[
		\sem{Hq;\; q := \ket{0}}\{\ket{0}\} = \{\ket{0}\bra{0} \hgate \ket{0}\} \cup \{\xgate \ket{1}\bra{1} \hgate \ket{0}\}  = \{\ket{0}\}
	\]
\end{example}

\begin{figure}
\begin{minipage}{0.52\textwidth}
\begin{align*}
	\tsem{\mathbf{skip}} &= \{\mathbf{skip}\}   \\
	\tsem{q := \ket{0}} &=  \{\mathbf{assume} \; P_q^0\}\cup \{\mathbf{assume} \; P_q^1;\; X q\} \\
	\tsem{U\qlist} &= \{U\qlist\} \\
	\tsem{\mathbf{meas} \; q} &= \{\mathbf{assume} \; P_q^0\}\cup\{\mathbf{assume} \; P_q^1\}
\end{align*}
\end{minipage}
\quad
\begin{minipage}{0.4\textwidth}
\begin{align*}
	\tsem{\mathbf{call} \; p(\qlist)} &= \tsem{p(\qlist)} \\
	\tsem{T_1; \; T_2 } &= \tsem{T_1}\compose \tsem{T_2} \\
	\tsem{\mathbf{if} \; \star\; \mathbf{then}\; T_1\; \mathbf{else} \; T_2} &= \tsem{T_1}\cup\tsem{T_2} \\
	\tsem{\mathbf{while} \; \star\; \mathbf{do} \; T} &= \cup_{k\geq 0}\tsem{T}^k
\end{align*}
\end{minipage}
\caption{Path collecting semantics for the non-deterministic quantum WHILE language.}\label{fig:csemantics}
\end{figure}

We extend $\supp{\cdot}$ to sets of quantum states in the obvious way. Given a set $s\in \powerset{\mathcal{C}_\mathcal{Q}}$ where $\mathcal{C}_\mathcal{Q} \defeq \F_2^{|\mathcal{Q}|}$ is the set of classical states on $\mathcal{Q}$ we can conversely take the linear span of quantum states $\{\ket{\vec x} \mid \vec{x} \in s\}$ as an element of $\powerset{\Hilb_{\mathcal{Q}}}$. We denote this operator simply as $\spn{s}\in\powerset{\Hilb_{\mathcal{Q}}}$. Note that $\supp{\cdot}$ and $\spn{\cdot}$ form a Galois connection between $(\powerset{\Hilb_{\mathcal{Q}}}, \subseteq)$ and $(\powerset{\mathcal{C}_\mathcal{Q}}, \subseteq)$:

\[(\powerset{\Hilb_{\mathcal{Q}}}, \subseteq) \galois{\mathrm{supp}}{\mathrm{span}} (\powerset{\mathcal{C}_\mathcal{Q}}, \subseteq) \]

A conceptual contribution of this paper is the fact that we may productively analyze a quantum program which includes both quantum and classical data flow (i.e. in and out of superposition) as a \emph{purely classical program} by abstracting its classical semantics $\mathcal{C}\sem{\cdot}:\powerset{\mathcal{C}_\mathcal{Q}}\rightarrow \powerset{\mathcal{C}_\mathcal{Q}}$.

\subsection{Quantum phase folding}

The \emph{phase folding} algorithm originated in \cite{amm14} from the \emph{phase polynomial} representation of \cxgate-dihedral operators. It was shown that any circuit over the \cxgate-dihedral gate set $\{\cxgate, \xgate, \rzgate(\theta) \mid \theta\in \mathbb{R}\}$ implements a unitary transformation which can be described as
\[
	U:\ket{\vec{x}} \mapsto e^{2\pi i f(\vec{x})}\ket{A\vec{x} + \vec{c}}
\]
where $(A,\vec{c}):\vec{x} \mapsto A\vec{x} + \vec{c}$ is an affine transformation over $\F_2$, and $f(\vec{x}) = \sum_{i}a_{i}f_i(\vec{x})\in \R$ where each $f_i:\F_2^n\rightarrow \F_2$ is a \emph{linear} function of $\vec x$. Note that a linear functional $f_i:\F_2^n\rightarrow \F_2$ is an element of the \emph{dual space} $(\F_2^n)^*$ and can hence be represented as a (row) vector. We use the two notions interchangeably. We call the linear functionals $\{f_i\}_i\subseteq (\F_2^n)^*$ for which $a_i\neq 0$ the \emph{support} of $f$. The utility of the phase polynomial representation for circuit optimization relies on two facts \cite{amm14}:
\begin{enumerate}
	\item for any \cxgate-dihedral circuit $U$, the canonical phase polynomial of $U$ can be efficiently computed and its support has size at most the number of $\rzgate$ gates in the circuit, and
	\item any unitary $U:\ket{\vec{x}} \mapsto e^{2\pi i f(\vec{x})}\ket{A\vec{x} + \vec{c}}$ can be efficiently implemented with exactly $|\supp{f}|$ $\rzgate$ gates, with parameters given by the Fourier coefficients $a_i$.
\end{enumerate}
Letting $\tau(U)$ denote the number of $\rzgate$ gates in $U$, the two facts above give a poly-time method of producing from $U$ a circuit $U'$ where $\tau(U')\leq \tau(U)$ -- notably, by computing $U$'s phase polynomial representation, then synthesizing a new circuit with at most $\supp{f}\leq \tau(U)$ $\rzgate$ gates.

\begin{example}\label{ex:one}
Consider the circuit, for which the intermediate states of the circuit as a function of an initial classical state $\ket{x,y}$ have been annotated, below:
\[
	\Qcircuit @C=1em @R=.2em @!R {
 		\lstick{x} & \ctrl{1} & \qw & \qw & \ctrl{1} & \qw & \targ & \push{x\oplus y}\qw & \gate{T^\dagger} & \targ & \rstick{x}\qw \\
 		\lstick{y} & \targ & \push{x\oplus y}\qw & \gate{T} & \targ & \push{y}\qw & \ctrl{-1} & \qw & \qw & \ctrl{-1} & \rstick{y}\qw
	}
\]
The effect of the first $\cxgate$ is to map $\ket{x,y}$ to $\ket{x,x\oplus y}$, at which point the $\tgate$ then applies a phase of $\omega$ conditional on the sum $x\oplus y$. After the third $\cxgate$ the state is now $\ket{x\oplus y, x}$ at which point the $\tgate^\dagger$ gate applies a phase of $\overline{\omega}$ conditional on the sum $x\oplus y$. It can be readily observed that for any $x,y\in\F_2$ the total phase accumulated by the circuit is $1$, as when $x\oplus y = 0$ no phase is accumulated, and when $x \oplus y = 1$ the phase accumulated is $\omega\overline{\omega}=1$.
\end{example}

The phase polynomial based optimization was extended in \cite{amm14} to a universal gate set by using path integrals. In particular, a circuit over the universal gate set $\{\hgate, \cxgate, \xgate, \rzgate(\theta)\}$ can be represented as the mapping
\[
	U:\ket{\vec{x}} \mapsto \sum_{\vec{y}\in\F_2^k}e^{2\pi i f(\vec{x},\vec{y})}(-1)^{Q(\vec{x},\vec{y})}\ket{A(\vec{x},\vec{y}) + \vec{c}}
\]
where $f$ and $A$ are as before, and $Q:\F_2^n\times\F_2^k\rightarrow \F_2$ is pure quadratic. While re-synthesis can be performed by splitting the above into \cxgate-dihedral sub-circuits, often re-synthesis is not only a bottleneck in efficiency, but may actually have the undesirable affect of \emph{increasing} the circuit cost in some other metric. For this reason, subsequent algorithms \cite{nrscm17,a19} dropped re-synthesis in favour of \emph{merging} existing $\rzgate$ gates provided they contribute to the same term of $f$. This also allowed $\rzgate$ gates with indeterminate parameters to be merged, as $\rzgate(\theta)\rzgate(\theta') = \rzgate(\theta + \theta')$.

\begin{example}\label{ex:had}
Consider the circuit below with intermediate states annotated again:
\[
	\Qcircuit @C=0.8em @R=.4em @!R{
 		\lstick{x} & \qw & \ctrl{1} & \qw & \targ & \push{y}\qw & \ctrl{1} & \qw & \qw 
 			& \qw & \ctrl{1} & \qw & \targ & \push{z}\qw & \ctrl{1} & \qw & \qw & \qw & \rstick{z}\qw \\
 		\lstick{y} & \gate{T} & \targ & \push{x\oplus y}\qw & \ctrl{-1} & \qw & \targ & \push{x}\qw & \gate{H}
 			& \push{z}\qw & \targ & \push{y\oplus z}\qw & \ctrl{-1} 
 			& \qw & \targ & \push{y}\qw & \gate{T} & \qw & \rstick{y}\qw
	}.
\]
The circuit implements the transformation $\ket{x,y} \mapsto \frac{1}{\sqrt{2}}\sum_{z\in\F_2} \omega^{2y}(-1)^{xz}\ket{z,y}$ where $z$ denotes to the branch taken (in superposition) by the Hadamard gate. Here we can see that the $T$ and $T^\dagger$ gates both contribute phases of $\omega^y$, giving a combined phase of $\omega\omega^{2y}=i^y$. Hence we can replace the two $\tgate$ gates with a single $\pgate$ gate at either location.
\end{example}

In \cite{a19}, one of the authors extended the optimization to include \emph{uninterpreted gates} --- gate symbols where a concrete semantics exists but is not known to the compiler. Such gates can be used to model gates from non-native gate sets, or \emph{opaque} gates and library calls as in the openQASM language \cite{qasm}. Intuitively, given an uninterpreted gate $U$ on $n$ qubits, $U$ can be assumed to map a classical state $\ket{x_1x_2\cdots x_n}$ to an arbitrary classical state $\ket{x_1'x_2'\cdots x_n'}$. It was shown that this interpretation is sound with respect to merging $\rzgate$ gates, as any rotation conditional on the primed outputs can't be commuted back through the uninterpreted gate. On the other hand, $(U \otimes I)\ket{x_1x_2\cdots x_n,y} = \ket{x_1'x_2'\cdots x_n', y}$ and so gates which involve qubits not modified by $U$ may commute and merge freely.

\begin{example}
Consider again the circuit above with the intermediate Hadamard gate replaced with an uninterpreted gate symbol $U$:
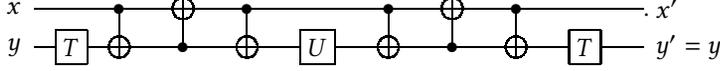
\[
	\Qcircuit @C=0.8em @R=.4em {
 		\lstick{x} & \qw & \ctrl{1} & \qw & \targ & \qw & \ctrl{1} & \qw & \qw 
 			& \qw & \ctrl{1} & \qw & \targ & \qw & \ctrl{1} & \qw & \qw & \qw & \rstick{x'}\qw \\
 		\lstick{y} & \gate{T} & \targ & \qw & \ctrl{-1} & \qw & \targ & \qw & \gate{U}
 			& \qw & \targ & \qw & \ctrl{-1} 
 			& \qw & \targ & \qw & \gate{T} & \qw & \rstick{y'=y}\qw
	}.
\]
While the circuit has no concrete semantics owing to the fact that $U$ is an indeterminate gate, the two $\tgate$ gates are able to be merged since regardless of the semantics of $U$, both phase gates apply if and only if $y=1$. The result is a total phase of $\omega^{2y}=i^y$, and hence can again be simulated with a single $\pgate$ gate at either location. Note that while the optimization of \cref{ex:had} is achievable by previous phase folding and related techniques \cite{amm14,nrscm17,zc19,kv20} which rely on \emph{exact} representations of the circuit semantics, \emph{these techniques are generally insufficient for the above optimization}.
\end{example}

In the context of program analysis, the mapping $U:\ket{x_1x_2\cdots x_n} \mapsto \ket{x_1'x_2'\cdots x_n'}$ asserts that \emph{no relations hold} between an input state and output state. Likewise, the $\cxgate$ mapping $\cxgate:\ket{x,y}\mapsto \ket{x, y\oplus x}$ asserts that the affine relations $x'=x$ and $y'=x\oplus y$ hold.  As the output of a quantum gate may in fact be a \emph{superposition} of classical states $\vec{x}'$, a relation on the pre- and post-state holds if it holds whenever $\bra{\vec x'}U\ket{\vec x} \neq 0$. This observation serves as the basis of our work.

\section{A relational approach to phase folding}\label{sec:ara}

In this section we formulate the phase folding optimization as a relational analysis computing (an approximation of) the affine relations between the classical inputs and outputs of a quantum circuit. We then use existing techniques \cite{karr76,elsar14} for affine relational analyses to extend this to non-deterministic quantum WHILE programs.

\subsection{From phase folding to subspaces}

In the phase folding algorithm, the (classical) support of the circuit $\ket{\vec x'}$ at any given point in the circuit is described as an affine function of the input variables $\vec x$ and intermediate variables $\vec y$. In particular, $x_i' = A_i[\vec{x}, \vec{y}]^T + c_i$. A $\rzgate(\theta)$ gate at this location then applies a phase of $e^{i\theta x_i'} = e^{i\theta (A_i[\vec{x}, \vec{y}]^T + c_i)} $, i.e. a phase of $e^{i\theta}$ conditional on $f_i(\vec x, \vec y) = A_i[\vec{x}, \vec{y}]^T  + c_i = 1$. Another phase gate which applies a phase conditional on $x_j' = A_j[\vec{x}, \vec{y}]^T  + c_j = 1$ contributes to the same term of the phase if and only if $A_i[\vec{x}, \vec{y}]^T = A_j[\vec{x}, \vec{y}]^T$ for all $\vec x, \vec y$. Note here that the affine factors of $x_i'$ and $x_j'$ may differ, in which case merging the phase gates results in an unobservable global phase.

Relations of the above form correspond \cite{karr76} to an \emph{affine subspace} over the variables (primed, unprimed, and in our case, intermediate), in the sense that the set of solutions $[\vec x', \vec x, \vec y]^T\in\F_2^{2n+k}$ to the relations $x_i' = A_i[\vec{x}, \vec{y}]^T + c_i$ form an affine subspace of $\F_2^{2n+k}$, where $n$ is the number of program variables and $k$ is the number of intermediate variables. This affine subspace can be represented as the kernel of a \emph{constraint matrix} $T\in\F_2^{n\times 2n+k+1}$, where the rows of $T$ encode the relations or \emph{constraints} $f_i(\vec{x'}, \vec{x}, \vec{y}) = c_i$ for linear functions $f_i$. In particular, the smallest affine subspace satisfying the relations $x_i' = A_i[\vec{x}, \vec{y}]^T + c_i$ can be encoded as the constraint matrix $T= \left[\begin{array}{c|c|c} I & A & \vec c \end{array}\right]$, noting that $T[\vec x', \vec x, \vec y, 1]^T = 0$ if and only if $x_i' = A_i[\vec{x}, \vec{y}]^T + c_i$ for every $i$. More generally, given a set of affine constraints $\{f_i\}_i^k$ (i.e. affine functions), we denote the smallest affine subspace satisfying each $f_i$ by $\langle f_1, f_2, \dots, f_k\rangle$ and encode it as the kernel of the constraint matrix with rows $f_i$. As $\F_2$ is a field, this subspace can further be uniquely represented by reducing the constraint matrix to reduced row echelon form.

Given two phases conditional on \emph{linear} functions $f_i,f_j\in(\F_2^{2n+k})^*$ of the primed, unprimed, and intermediate variables, $f_i$ and $f_j$ correspond to row vectors over $\F_2^{2n+k}$ and $f_i(\vec x',\vec x, \vec y) = f_j(\vec x',\vec x, \vec y)$ for all $\vec z\in \ker T$ if and only if as row vectors, $f_i + f_j$ is in the row space of $T$ up to an (irrelevant) affine factor. Moreover, the restriction of $f_i$ and $f_j$ to $\ker T$ can be uniquely canonicalized by fully row reducing them with respect to $T$. Hence, to merge all phases conditional on linear functions over $\vec x', \vec x$, and $\vec y$, we need only row reduce each one with respect to $T$ and collect phases on identical conditions.

\begin{example}
Consider the circuit below, reproduced from \cref{ex:one}, with intermediate variables $t_1$, $t_2$ at distinguished points.
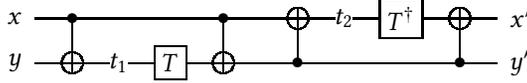
\[
	\Qcircuit @C=1em @R=.2em @!R {
 		\lstick{x} & \ctrl{1} & \qw & \qw & \ctrl{1} & \qw & \targ & \push{t_2}\qw & \gate{T^\dagger} & \targ & \rstick{x'}\qw \\
 		\lstick{y} & \targ & \push{t_1}\qw & \gate{T} & \targ & \qw & \ctrl{-1} & \qw & \qw & \ctrl{-1} & \rstick{y'}\qw
	}
\]
The non-zero transitions of this circuit lie in the subspace  $\langle x'=x, y'=y, t_1=x\oplus y, t_2=x\oplus y\rangle $ which we can encode as the constraint matrix on the left below, reduced to row echelon form on the right:
\[
\begin{NiceArray}{lcc|cc|cc|c}[first-row]
\text{constraint} & x' & y' & x & y & t_1 & t_2 & c \\
x'=x  & 1 & 0 & 1 & 0 & 0 & 0 & 0 \\
y'=y & 0 & 1 & 0 & 1 & 0 & 0 & 0 \\
t_1=x\oplus y  & 0 & 0 & 1 & 1 & 1 & 0 & 0 \\
t_2=x\oplus y  & 0 & 0 & 1 & 1 & 0 & 1 & 0
\CodeAfter \SubMatrix[{1-2}{4-8}]
\end{NiceArray}\quad \longrightarrow\quad  
\begin{NiceArray}{lcc|cc|cc|c}[first-row]
\text{constraint} & x' & y' & x & y & t_1 & t_2 & c \\
x'=y\oplus t_2  & 1 & 0 & 0 & 1 & 0 & 1 & 0 \\
y'=y  & 0 & 1 & 0 & 1 & 0 & 0 & 0 \\
x=y\oplus t_2  & 0 & 0 & 1 & 1 & 0 & 1 & 0 \\
t_1=t_2 & 0 & 0 & 0 & 0 & 1 & 1 & 0
\CodeAfter \SubMatrix[{1-2}{4-8}]
\end{NiceArray}
\]
The two phases of $\omega$ and $\overline{\omega}$ are conditional on $t_1$ and $t_2$ respectively, corresponding to the linear functionals
\[
f = \begin{bNiceArray}{cc|cc|cc}[first-row,first-col]
& x' & y' & x & y & t_1 & t_2 \\
 & 0 & 0 & 0 & 0 & 1 & 0
\end{bNiceArray} \qquad \text{and} \qquad 
g=\begin{bNiceArray}{cc|cc|cc}[first-row,first-col]
& x' & y' & x & y & t_1 & t_2 \\
 & 0 & 0 & 0 & 0 & 0 & 1
\end{bNiceArray}
\]
Row reducing both modulo (the linear part of) the reduced constraint matrix above yields the same linear functional ($t_2$), hence both phases contribute to the same term and can be canceled.
\end{example}

\subsection{Phase folding as an affine relation analysis}

The previous observations are not new in either the classical or in the quantum context, but the shift in perspective allows one to re-frame the phase-folding optimization as computing an affine subspace $\subspace\subseteq \F_2^{2n}$ corresponding to the classical transitions of a circuit, for which many domains have previously been devised \cite{karr76,ms04,ks08,elsar14}. Before discussing concrete domains and their use in the phase folding optimization, we first formalize phase folding as an affine relation analysis. 

Using the terminology of \cite{elsar14}, we say that $\subspace[\vec X'\sep \vec X\sep \vec Y]\subseteq \F_2^{2n+k}$ is an affine subspace in three \emph{vocabularies} --- $n$ primed variables $\vec X'$ representing the post-state, $n$ unprimed variables $\vec X$ representing the pre-state, and $k$ intermediate variables $\vec Y$. The definition below formalizes the notion of such a subspace as a sound abstraction of the classical transitions of a control-flow path.

\begin{definition}[sound \& precise]
We say that an affine subspace $\subspace[\vec X'\sep \vec X\sep \vec Y]$ is a \emph{sound abstraction} of a control-flow path $\pi\in\Sigma^*$ if for any $\vec x', \vec x \in \F_2^{n}$
\[
\vec x' \in \supp {\sem{\pi}\ket{\vec x}}\implies \exists \vec y. (\vec x', \vec x, \vec y)\in \subspace
\]
We say that $\subspace$ is a \emph{precise} abstraction if the reverse implication also holds.
\end{definition}

The condition above that $ \exists \vec y. (\vec x', \vec x, \vec y)\in \subspace$ states that $(\vec x', \vec x)$ is contained in the affine subspace $\subspace'$ of $\F_2^{n}\times\F_2^{n}$ with the coordinates corresponding to $\vec Y$ projected out. We denote this subspace $\exists \vec Y. \subspace$ and observe that $\subspace[\vec X'\sep \vec X\sep \vec Y]$ is a sound (resp. precise) abstraction of $\pi$ if and only if $\exists \vec Y.\subspace$ is. We denote a subspace in two vocabularies $\vec X'$, $\vec X$ --- for instance, a three-vocabulary subspace with intermediates projected out --- by $\subspace[\vec X'\sep \vec X]$. 

Intuitively, a subspace $\subspace[\vec X'\sep \vec X]$ which soundly approximates a program path $\pi$ over-approximates the set of non-zero transitions $\bra{\vec x}\sem{\pi}\ket{\vec x}$ and can be viewed as a \emph{state transition formula} \cite{krc21}. Note that if $\pi$ is infeasible, then $\supp{\sem{\pi}\ket{\vec x}} = \emptyset$ for any $\vec x$, hence the special empty subset $\bot = \emptyset$ is a sound abstraction of $\pi$. We can compose two state transitions $\subspace[\vec X'\sep \vec X]$, $\subspace'[\vec X'\sep \vec X]$ by taking their relational composition 
\[(\vec x', \vec x)\in \subspace \compose \subspace' \iff \exists \vec x''\in\F_2^n. (\vec x'', \vec x)\in \subspace \land (\vec x', \vec x'')\in \subspace'\]
which is again an affine subspace corresponding to the intersection of $\subspace[\vec X''\sep \vec X]$ and $\subspace'[\vec X'\sep \vec X'']$ taken as subspaces of the three-vocabulary space $\F_2^{3n}$ followed by a projection of $\vec X''$. We denote the projection of $\vec X''$ by $\exists \vec X''.\subspace$, and define \[\subspace \compose \subspace' \defeq \exists X''.\subspace[\vec X''\sep \vec X] \cap \subspace'[\vec X'\sep \vec X''].\]

\begin{proposition}
	If $\subspace$ and $\subspace'$ are sound abstractions of $\pi$ and $\pi'$, then $\subspace\compose \subspace'$ is sound for $\pi\compose \pi'$.
\end{proposition}
\begin{proof}
	Follows from linearity of $\sem{\pi\compose \pi'} = \sem{\pi'}\sem{\pi}$.
\end{proof}

\begin{remark}
In contrast to classical contexts, composition is \emph{not precise} due to the effects of interference. Notably, the only sound abstraction of $\hgate$ is $\top = \F_2^2$ since we have non-zero transitions between every $\ket{x}$ and $\ket{x'}$. Clearly $\top\compose\top = \top$. On the other hand, $\vec x' \in \supp {\sem{\hgate\hgate}\ket{\vec x}}\implies x'=x$, and so a sound abstraction of $\hgate\hgate$ is the $1$-dimensional affine subspace of $\F_2^2$ satisfying $x'=x$.

We can't generally expect to compute precise subspace abstractions in polynomial time, even for circuits which implement affine transitions, as a polynomial-time solution to the problem of \emph{strong simulation} --- computing $\bra{\vec x'}U\ket{x}$ given a circuit $U$ --- would imply $\mathbf{BQP}=\mathbf{P}$ \cite{dhmhno05}. While the problem of determining a precise abstraction of a circuit as an affine transformation is slightly weaker in that it only tells us when $\bra{\vec x'}U\ket{x}\neq 0$, it can be noted that this suffices to efficiently simulate several \emph{deterministic} quantum algorithms which exhibit exponential query complexity separation between quantum and classical algorithms \cite{bv97,r08}.
\end{remark}

The following proposition establishes a basis for merging phase gates over affine transition relations. In particular, if $x_j' = x_i$ in every non-zero transition of a control-flow path, then a phase gate on qubit $i$ in the prefix may be merged with a phase gate on qubit $j$ in the suffix.

\begin{proposition}\label{prop:pf}
Let $\subspace[\vec X'\sep \vec X]$ be a sound abstraction of a path $\pi$ and suppose that for every $(\vec x',\vec x)\in \subspace$, $x_j' = x_i$.
Then $\sem{\rzgate(\theta)_{q_i}\compose \pi\compose \rzgate(\phi)_{q_j}} = \sem{\rzgate(\theta+\phi)_{q_i}\compose \pi\compose \mathbf{skip}} = \sem{\mathbf{skip}\compose  \pi\compose \rzgate(\theta+\phi)_{q_j}}.$
\end{proposition}
\begin{proof}
By computation, noting that $\vec{x}'\in \supp{\sem{\pi}\ket{\vec x}}$  only if $(\vec x',\vec x)\in \subspace$ and hence $x_j'=x_i$:
\begin{align*}
		\sem{\rzgate(\theta)_{q_i}\compose \pi\compose \rzgate(\phi)_{q_j}}\ket{\vec{x}}
		&= e^{i\theta x_i}\left(\rzgate(\phi)_{q_j}\sem{\pi}\ket{\vec x}\right) \\ 
		&= e^{i\theta x_{i}} \left(\rzgate(\phi)_{q_j}\sum\nolimits_{\vec{x}'\in\supp{\sem{\pi}\ket{\vec x}}} \alpha_{\vec{x}'}\ket{\vec{x}'}\right) \\
		&= e^{i\theta x_{i}} \left(\sum\nolimits_{\vec{x}'\in\supp{\sem{\pi}\ket{\vec x}}} e^{i\phi x'_j}\alpha_{\vec{x}'}\ket{\vec{x}'}\right) \\
		&= e^{i(\theta + \phi)x_{i}}\sem{\pi}\ket{\vec x}
\end{align*}
The claim then follows by linearity.
\end{proof}

Note that \cref{prop:pf} gives a method of merging phases applied \emph{along} a path. In particular, if we view $x_i$ and $x'_j$ as linear functionals $f_i(\vec x ', \vec x)= x_i$, $f_j(\vec x', \vec x) = x'_j$, respectively, then we may represent the \emph{condition} $f\in(\F_2^{2n})^*$ of each phase gate via a canonical representative (if it exists) on the subspace $\subspace[\vec X'\sep \vec X]$. If two conditions $f_i + f_j \in \ker \subspace[\vec X'\sep \vec X]$ where $\ker\subspace$ is the subspace of $(\F_2^{2n})^*$ which annihilates $\subspace$, then by definition $f_i(\vec x', \vec x) = f_j(\vec x', \vec x)$ for all $(\vec x', \vec x)\in\subspace$.

\subsection{Extending to quantum programs}

The developments of the previous section allow the phase folding algorithm to be easily extended to non-deterministic quantum WHILE programs. Given a set of paths $\Pi$, a two-vocabulary subspace $\subspace\subseteq \F_2^{2n}$ is a sound abstraction of $\Pi$ if and only if $\subspace$ is a sound abstraction of every $\pi\in\Pi$. Correctness of the phase folding condition, i.e. \cref{prop:pf}, carries over to sets of paths in this way, as phase gates may be merged if and only if they can be merged along every path. 

Given a concrete domain of subspaces of $\F_2^{2n}$, we may compute abstractions of sets of paths from abstractions of individual paths in the standard way by taking the \emph{join}, corresponding to the \emph{affine hull} of subspaces. In particular, if $\subspace_1$ and $\subspace_2$ are sound abstractions of $\pi_1$ and $\pi_2$, it is easy to see that the smallest subspace which soundly abstracts both is the affine hull of $\subspace_1$ and $\subspace_2$, denoted $\subspace_1\sqcup \subspace_2$ where
\[
	\subspace_1\sqcup \subspace_2 \defeq \spn{\{\vec x \mid \vec x \in  \subspace_1\cup \subspace_2 \}}
\]
Moreover, as noted by \cite{karr76}, since the subspaces of $\F_2^{2n}$ have finite and bounded dimension, there can be no infinite strictly ascending chain $\subspace_1 \subsetneq \subspace_2 \subsetneq \cdots$
and so the join of infinitely many paths stabilizes in finitely many ($O(n)$) steps. As is customary \cite{krc21} we define the \emph{Kleene closure} $\subspace^\kleene$ of a two-vocabulary subspace $\subspace[\vec X'\sep \vec X]$ to be the limit of the sequence $\{\subspace_i\}_{i=0}^\infty$ where
\[
	\subspace_0 = \bot=\{0\} \qquad\qquad  \subspace_{i+1} = \subspace_i \sqcup (\subspace_i \compose \subspace)
\]
which satisfies $\subspace_{i}\subseteq \subspace_{i+1}$ for all $i\geq 0$ and hence stabilizes in at most $2n$ steps.

We now have all the pieces necessary to define an affine relation analysis for quantum WHILE programs. Let
\[
	(\subspace(\F_2^{2n}), \subseteq, \sqcup, \sqcap, \bot, \top)
\]
be some domain of affine transition relations where $\subspace(\F_2^{2n})$ is the set of affine subspaces of $\F_2^{2n}$ with a special element $\bot$ for the empty subset, $\sqcup$ is the union of affine subspaces, $\sqcap$ is subspace intersection, and $\top=\F_2^{2n}$. Additionally, let $\subspace(\F_2^{2n})$ be equipped with an additional composition operator $\compose$. \Cref{fig:ara} gives an abstract semantics $\abssem{T}\in\subspace(\F_2^{2n})$ for computing a sound approximation of a quantum WHILE program over typical gates as an affine transition function. Transition relations for gates and projections are given as a function of the qubits they act on --- we naturally extend the transition relation of a gate on qubits $\qlist$ to a transition relation on $\mathcal{Q}$ by setting the pre- and post-state for any qubit in $\mathcal{Q}-\qlist$ to be equal.

\begin{figure}

\begin{minipage}{0.4\textwidth}
\begin{align*}
	\abssem{\xgate} &= \langle x_q' \oplus 1\oplus x_q\rangle  \\
	\abssem{\rzgate(\theta)_q}&= \langle x_q' \oplus x_q\rangle \\
	\abssem{\hgate_q} &= \top_q \\
	\abssem{\cxgate_{q_1q_2}} &= \langle x_{q_1}' \oplus x_{q_1},  x_{q_2}' \oplus x_{q_1}\oplus x_{q_2}\rangle \\
	\abssem{P^b_q} &= \langle x_q' \oplus x_q,  x_q \oplus b\rangle \\
	\abssem{U\qlist} &= \abssem{U}[X_{\qlist}'\sep X_{\qlist}]
\end{align*}
\end{minipage}
\begin{minipage}{0.55\textwidth}
\begin{align*}
	\abssem{q:= \ket{0}} &= \langle x_q'\rangle \\
	\abssem{\mathbf{meas}\; q} &=  \langle x_q' \oplus x_q\rangle \\
	\abssem{\mathbf{call} \; p(\qlist)} &= \abssem{p(\qlist)} \\
	\abssem{T_1; \; T_2 } &= \abssem{T_1} \compose \abssem{T_2} \\
          \abssem{\mathbf{if} \; \star\; \mathbf{then}\; T_1\; \mathbf{else} \; T_2} &= \abssem{T_1}\sqcup \abssem{T_2} \\
	\abssem{\mathbf{while} \; \star\; \mathbf{do} \; T} &= \abssem{T}^\kleene
\end{align*}
\end{minipage}
\caption{Affine relation analysis for quantum WHILE programs}\label{fig:ara}
\end{figure}

The analysis of \cref{fig:ara} is defined without respect to a particular implementation of the subspace domain. In principle any \emph{relational} domain for affine subspaces suffices, the key element of relational in this case being the ability to compose relations and hence abstract affine \emph{transfer} functions rather than sets of states. The $\mathbf{KS}$ domain of \cite{elsar14} is one such concrete implementation which suffices for the purpose of quantum ARA. In the next section we develop a concrete domain for performing phase folding with an affine relation analysis which is based on the $\mathbf{KS}$ domain.

\begin{theorem}\label{thm:sound}
Let $T$ be a quantum WHILE program. Then $\abssem{T}$ is a sound abstraction of $\tsem{T}$.
\end{theorem}
\begin{proof}
It can be readily observed that the abstractions of basic gates are all sound (and precise), and likewise that the rules for unitary gate application, composition, non-deterministic choice and iteration are sound. It remains to show that the transition relations for reset and measurement are sound. For measurement, we note that 
$\tsem{\mathbf{meas}\; q} = \{\mathbf{assume}\; P_q^0\} \cup \{\mathbf{assume}\; P_q^1\}$, and hence a sound abstraction of $\mathbf{meas}\; q$ is $\abssem{P^0}\sqcup \abssem{P^1} = \langle x' \oplus x,  x\rangle\sqcup\langle x' \oplus x,  x\oplus 1\rangle = \langle x' \oplus x\rangle.$
Likewise, for reset we have $\tsem{q:= \ket{0}} = \{\mathbf{assume}\; P_q^0\} \cup \{\mathbf{assume}\; P_q^1\compose Xq\}$, which is soundly abstracted by the join
$\abssem{P^0}\sqcup \abssem{XP^1} = \langle x' \oplus x,  x\rangle\sqcup\langle x',  x\oplus 1\rangle = \langle x'\rangle.$
\end{proof}

\subsection{Forward ARA with summarization}\label{sec:fara}

We now give a concrete algorithm for phase folding with respect to affine transition relations over quantum WHILE programs. We use the $\mathbf{KS}$ domain of \cite{ks08,elsar14} for the affine relational analysis, tailored to the pecularities of the phase folding context. Notably,
\begin{enumerate}
	\item programs are \emph{large}: most quantum programs include large sections of straightline code involving thousands, if not millions, of gates. To handle practical examples of quantum circuits our analysis should scale to $m=10^3$--$10^6$ gates.
	\item Projection is \emph{expensive}: for a circuit involving $m$ gates, there are $O(m)$ phase terms which need to be normalized when projecting out a variable.
	\item Variables are \emph{inexpensive}: as our variables are binary, we can implement fast linear algebra over many variables by storing (row) vectors as bitvectors.
\end{enumerate}
To this end, our analysis is designed to avoid projection as much as possible, and to avoid whenever possible applying projection to phase conditions. This amounts to performing a \emph{forward analysis} which maintains the current state as an affine subspace over the pre-state and intermediate variables, together with \emph{summarization} of loops and branches using $\mathbf{KS}$ domain operations.

We first recall the definition of the $\mathbf{KS}$ domain from \cite{elsar14}. An element of the $\mathbf{KS}$ domain in vocabularies $\vec X', \vec X ,\vec Y$, denoted $\mathbf{KS}[\vec X'\sep \vec X\sep \vec Y]$ where $|\vec X|=n$ and $|\vec Y|=k$, is a matrix $A\in\F_2^{m\times 2n+k+1}$ in reduced echelon form with all-zero rows dropped. An element of the domain represents an affine $3$-vocabulary subspace $\subspace[\vec X'\sep \vec X\sep \vec Y]$ of $\F_2^{2n+k}$ as the (affine) kernel of $A$, $\ker A$, and its concretization is equal to this subspace. The rows of $A$ are viewed as affine constraints $A_i(\vec x', \vec x, \vec y) + c_i = 0$ where the first $n$ columns correspond to variables $\vec X'$, followed by the variables $\vec X$ and then the variables $\vec Y$. A special $1$-row element, $\bot = \begin{bmatrix} 0 & 0 & 0 \mid  1\end{bmatrix}$ is reserved to represent the empty subspace, and whenever $A$ contains such a row which is necessarily the case when $\ker A = \emptyset$, $A$ is automatically reduced to $\bot$. The top element $\top$ is the empty matrix corresponding uniquely to $\ker \top = \F_2^{2n+k}$.

Projection of $\vec X'$ for an element $A$ of $\mathbf{KS}[\vec X'\sep \vec X\sep \vec Y]$ is implemented by removing any row which is non-zero in one of the columns of $\vec X'$. Other variable sets may be projected out by re-ordering the columns to place the variables to be projected in the left-most columns, writing in row-echelon with the new variable order, and then projecting out those columns. The remaining domain operations can be defined via projection as below, where $A = \left[\begin{array}{ccc|c}
 A_{\text{Post}} & A_{\text{Post}} & A_{\text{Aux}} & \vec c
\end{array}\right]$, $B=\left[\begin{array}{ccc|c}
 B_{\text{Post}} & B_{\text{Post}} & B_{\text{Aux}} & \vec d
\end{array}\right]$ are elements of the $\mathbf{KS}$ domain and $\exists.$ denotes the projection of the left-most block:
\begin{align*}
	A\sqcap B = \begin{bmatrix} A \\ B \end{bmatrix} \qquad
	A\sqcup B = \exists. \left[\begin{array}{c|c}
 A & A \\
 B & 0 
\end{array}\right] \qquad
	A\compose B = 	\exists.\left[\begin{array}{c|cccc|c}
 A_{\text{Post}} & 0 & A_{\text{Pre}} & A_{\text{Aux}} & 0 & \vec c \\
 B_{\text{Pre}} & B_{\text{Post}} & 0 & 0 & B_{\text{Aux}} & \vec d \\
\end{array}\right]
\end{align*}

We define one additional operation on the $\mathbf{KS}$ domain, \[\mathrm{reduce}:\mathbf{KS}[\vec X'\sep \vec X\sep \vec Y]\times \dual{\F_2^{2n + k}} \rightarrow\dual{\F_2^{2n + k}}\] where $\dual{\F_2^{2n + k}}$ denotes the dual space of $\F_2^{2n + k}$. The operation $\mathrm{reduce}$ canonicalizes a linear functional $f\in\dual{\F_2^{2n + k}}$ with respect to an element $A$ of $\mathbf{KS}[\vec X'\sep \vec X\sep \vec Y]$ by row reducing the matrix $\left[\begin{smallmatrix} A \\ f \end{smallmatrix}\right]\rightarrow \left[\begin{smallmatrix} A \\ f' \end{smallmatrix}\right]$ and returning $f'$. Note that $f_1(\vec x', \vec x, \vec y) = f_2(\vec x', \vec x, \vec y)$ for all $(\vec x', \vec x, \vec y)\in\ker A$ if and only if $\mathrm{reduce}(A, f_1) = \mathrm{reduce}(A, f_2)$, since the reduction of either gives a unique representation of the kernel of $\left[\begin{smallmatrix} A \\ f_1 \end{smallmatrix}\right]$ and $\left[\begin{smallmatrix} A \\ f_2 \end{smallmatrix}\right]$, respectively. Note that in contrast to \cite{elsar14}, we write $\mathbf{KS}$ elements with the post-state first so that ordinary Gaussian elimination will eliminate those first, as our goal is to canonicalize over the pre- and intermediate- states. In practice, the pre- and intermediate-states are separable in our analysis, so their order does not matter. 

We can now describe the phase folding analysis. We denote the set of program locations by $\mathrm{Loc}$. A \emph{condition} $\mathrm{Cond}\defeq\dual{\F_2^{2n + k}}\cup \{\bot\}$ is either $\bot$ or $f\in\dual{\F_2^{2n + k}}$, and a \emph{phase term} is a pair $f(, L)$ of a condition $f$ and a subset of program locations $L$. We define
\[
	\mathbf{PF}_{\mathrm{Aff}} = \powerset{\mathrm{Cond}\times \powerset{\mathrm{Loc}}}\times \mathbf{KS}[\vec X'\sep \vec X\sep \vec Y].
\]
The intuition behind an element of $\mathbf{PF}_{\mathrm{Aff}}$ is that it represents as program $T$ as an affine subspace on the pre-, post-, and intermediate-states, along with a partition of the $\rzgate$ gates of $T$ into disjoint sets, each of which has a representative expressing the condition as a linear function of the pre-, post-, and intermediate-states. If no such function exists, the set has representative $\bot$.

The algorithm proceeds in an iterative fashion for basic blocks. The transfer functions $\mathbf{PF}_{\mathrm{Aff}}\sem{T}:\mathbf{PF}_{\mathrm{Aff}}\rightarrow \mathbf{PF}_{\mathrm{Aff}}$ are given in \cref{fig:fara}. The forward analysis maintains an element $(t, A)\in\mathbf{PF}_{\mathrm{Aff}}$ where $A=\begin{bmatrix} I & A_{\text{Pre}} & A_{\text{Aux}} \mid \vec c \end{bmatrix}$ so that $\vec x'=A_{\text{Pre}}\vec x + A_{\text{Aux}}\vec y + \vec c$. When a phase gate is applied to qubit $i$ at location $\ell$, $\ell$ is added to $t$ with condition $f = (A^{\text{Pre}+\text{Post}})_i$ given by the $i$th row of $A$ with the (irrelevant) affine factor dropped. If another phase term has the same (non-$\bot$) representative function $f$, the phase terms are merged by taking the union of their locations (denoted $\uplus$ in \cref{fig:fara}). Gates with transitions for which some $x_i'$ has no representative in terms of $X$ and $Y$, notably the $\hgate$ gate, add a new intermediate variable $Y\cup\{y_{k+1}\}$ and are modeled with the relation $x_i' = y_{k+1}$ to maintain $A$ in the above form. The initial value is defined as $(\emptyset, I)$.

\begin{figure}
\begin{minipage}[t]{0.6\textwidth}
\begin{align*}
	&\mathbf{PF}_{\mathrm{Aff}}\sem{\hgate_q}(t, A) = (t,A[x_q' \gets y_{k+1} \oplus 1]) \\
	&\mathbf{PF}_{\mathrm{Aff}}\sem{\cxgate_{q_1q_2}}(t, A) = (t,A[x_{q_2}' \gets x_{q_1} \oplus x_{q_2}]) \\
	&\mathbf{PF}_{\mathrm{Aff}}\sem{\rzgate(\theta)_q^{\ell}}(t, A) = (t\uplus ((A^{\text{Pre}+\text{Post}})_i, \{\ell\}), A)
\end{align*}
\end{minipage}
\begin{minipage}[t]{0.38\textwidth}
\begin{align*}
	&\mathbf{PF}_{\mathrm{Aff}}\sem{\xgate_q}(t, A) = (t,A[x_q' \gets x_q \oplus 1])  \\
	&\mathbf{PF}_{\mathrm{Aff}}\sem{q:= \ket{0}}(t, A) = (t,A[x_q' \gets 0])  \\
	&\mathbf{PF}_{\mathrm{Aff}}\sem{\mathbf{meas}\; q}(t, A) = (t,A)
\end{align*}
\end{minipage}
\[
	t_1\uplus t_2 = \{(\bot, L) \mid (\bot, L)\in t_1 \cup t_2\} \cup \{(f, L_1 \cup L_2) \mid (f,L_1)\in t_1 \land (f,L_2)\in t_2\}
\]
\caption{Forward analysis of basic blocks.}\label{fig:fara}
\end{figure}

\paragraph{Computing summaries}

To deal with choice, iteration, and procedure calls, the algorithm first analyzes the body or bodies and then produces a \emph{summary}. \Cref{fig:summary} gives our summarization rules, where $(t_1,A_1)\sqcup (t_2,A_2) \defeq (t_1\cup t_2, A_1\sqcup A_2)$ and $(t,A)^\kleene \defeq (t, A^\kleene)$. We distinguish summaries from iterative values of $\mathbf{PF}_{\mathrm{Aff}}$ as the summary generally won't be of a form where $\vec x'$ has an explicit representation in terms of $\vec x$ and $\vec y$ --- we say such relations are \emph{solvable}. For instance, at the end of a loop $\vec x'$ may depend on a temporary value initialized in the loop body, which must then be projected away to compute the Kleene closure. Moreover, phase gates applied inside a loop, conditional, or procedure can't cross the control-flow boundaries even if they apply on the same condition as some phase gate outside the loop. The exception is phase gates conditional on the zero-everywhere function, $\vec{0}^T$, which can safely be eliminated --- to identify these cases, the summary preserves any partitions whose condition is \emph{nullable} when composed with a pre-condition, and maps the representative of every other partition to $\bot$.

\begin{figure}
\begin{align*}
	&\mathbf{PF}_{\mathrm{Sum}}\sem{\mathbf{if} \; \star\; \mathbf{then}\; T_1\; \mathbf{else} \; T_2} = \mathbf{PF}_{\mathrm{Aff}}\sem{T_1}(\emptyset, I)\sqcup \mathbf{PF}_{\mathrm{Sum}}\sem{T_2}(\emptyset, I) \\
	&\mathbf{PF}_{\mathrm{Sum}}\sem{\mathbf{while} \; \star\; \mathbf{do} \; T} = \mathbf{PF}_{\mathrm{Aff}}\sem{T}(\emptyset, I)^\kleene \\
	&\mathbf{PF}_{\mathrm{Sum}}\sem{\mathbf{call} \; p(\qlist)} = (\{(\bot, L) \mid (-,L)\in t\}, A[X_{\qlist}', X_{\qlist}]) \qquad\qquad \text{where} (t, A) = \mathbf{PF}_{\mathrm{Aff}}\sem{T}(\emptyset, I)
\end{align*}
\caption{Summary computations for choice, iteration, and procedure calls.}\label{fig:summary}
\end{figure}

\begin{figure}
\begin{align*}
	\mathbf{PF}_{\mathrm{ff}}(t,A)(t',A') = (t\uplus t'_\text{null}, A\compose_{\mathrm{ff}} A'), \; t'_{\text{null}} = \left\{ (r , L) \mid (f, L)\in t' \land r = \begin{cases} 0 &\!\!\text{if}\; \text{reduce}(A, f) = 0 \\ \bot &\!\!\text{otherwise}\end{cases}\right\}
\end{align*}
\caption{Summary application.}\label{fig:apply}
\end{figure}

\paragraph{Applying summaries}

To apply summaries, we must compose a solvable relation $A$ with some potentially non-solvable relation $A'$ into a once again solvable relation. To do so, we add an extra vocabulary of intermediate variables $\vec Y'$ and assert the equality between the post-state $\vec X'$ and $\vec Y'$. Explicitly, we define the summary composition operator $\compose_{\mathrm{ff}}$ as
\[
	A\compose_{\mathrm{ff}} A' = A\compose A' \sqcap \langle \vec X' = \vec Y'\rangle =\exists.\left[\begin{array}{c|cccc|c}
 A_{\text{Post}} & 0 & A_{\text{Pre}} & A_{\text{Aux}} & 0 & \vec c \\
 A'_{\text{Pre}} & A'_{\text{Post}} & 0 & 0 & 0 & \vec d \\
 0 & I & 0 & 0 & I & \vec{0} \\
\end{array}\right]
\]

In many practical contexts such as in Grover's algorithm, an ancilla is often initialized in the $\ket{0}$ state prior to entering a loop, with the invariant that the ancilla is returned to the initial state after each iteration. Such situations can be used to perform optimizations \emph{inside} the loop body, particularly if the assumption that the ancilla is in the $\ket{0}$ leads to a phase condition becoming null (i.e. $f = 0$). To make use of these optimizations, when composing a domain with a summary for a block we check whether the pre-state defined by the domain element implies any of the phase gates in the summarized block are nullable. To do so, we reduce the phase terms inside the summarized block with the pre-state and record the phases whose conditions have become null, as in \cref{fig:apply}. 

\begin{theorem}\label{thm:corr}
	Let  $\mathbf{PF}_{\mathrm{Aff}}\sem{T} = (t, A)$. Then for every $(f, L)\in t$, the phase gates at locations $\ell\in L$ in $T$ can be safely merged into a single gate, or eliminated if $f=0$. Moreover, $\mathbf{PF}_{\mathrm{Aff}}\sem{T}$ can be computed in time $O(n^3m + nm^2\log m)$ where $n$ is the number of qubits and $m$ is the number of program locations.
\end{theorem}
\begin{proof}
Correctness follows from \cref{thm:sound} and \cref{prop:pf}. For the complexity we assume that bit vectors are stored as packed integers, hence operations on bit vectors take time $O(1)$. At each step in the analysis, $t$ has at most $m$ terms, and $A$ is an $n\times 2n+k+1$ matrix represented as a list of $n$ bit vectors. Row operations corresponding to gate applications hence take $O(1)$ time, adding to $t$ takes time $O(\log m)$, while Gaussian elimination for computing joins takes $O(n^2)$ time. Loop summaries stabilize in at most $n$ iterations, since the initial dimension of the loop transition subspace is $n$, giving a total cost of $O(n^3)$ to compute a loop summary. Finally, applying a summary involves a Gaussian elimination at cost $O(n^2)$ as well as reduction of all terms of $t$, each at a cost of $O(n)$ for a total cost of $O(nm\log m)$ to reduce and merge all terms of $t$. The resulting complexity is hence bounded by $O(n^3m+ nm^2\log m)$.
\end{proof}

\section{Non-linear phase folding}\label{sec:pra}

Re-casting the phase folding optimization as a relational analysis not only allows the original \emph{affine} analysis to be generalized to quantum programs, but also allows the optimization to be easily generalized to \emph{non-linear} analyses. Such analyses can be used to capture instances of phase folding where the non-linear Toffoli gate $\ccxgate$ gate is used like in the motivating example, as well as situations where affine transition relations give rise to complicated non-linear loop invariants. Note that by non-linear here we mean that the Toffoli gate, while linear as a unitary operator over $\C^{2^3}$, computes a non-linear permutation of $\F_2^3$ expressed as the relation $\ccxgate:\ket{x,y,z} \mapsto \ket{x, y, z\oplus xy}$. In this section we present one such analysis based on abstracting classical transitions via \emph{polynomial ideals} which suffices to achieve both optimizations. While the connection between polynomial ideals and loop invariants has been well-established \cite{rk04,ssm04,ck24}, the simplistic nature of the underlying finite field $\F_2$ allows us to do away with most of the complications involved in similar classical program analyses. Moreover, our techniques are illustrative of the flexibility of our algorithm in making use of classical techniques for relational analysis.

%%%%%%%%%%%%%%%%%%%%%%%%

\subsection{Polynomial transition ideals}

We first recall some basic details of algebraic geometry \cite{cox}. An \emph{affine variety} over $\F_2^n$ is the set of simultaneous solutions of a set $S\subseteq \F_2[X_1\dots, X_n]$ of polynomial equations:
\[
	V=\variety(S) = \{ \vec x \in \F_2^n \mid f(\vec x) = 0\; \forall f\in S\}.
\]
It is a basic fact that the polynomials which vanish on the variety $V$ form an ideal $I$ in $\F_2[\vec X]$,
\[
	I=\ideal(V) = \{ f\in \F_2[\vec X] \mid f(\vec x) = 0\; \forall \vec x \in V\}.
\]
Given an algebraic variety $V$ over $\vec X$, the polynomial ideal $\ideal(V)$ can be viewed as the set of \emph{polynomial relations on $\vec X$} which are satisfied by $V$, in the same way that an affine \emph{subspace} can be viewed as a set of affine (degree 1) \emph{relations} satisfied by elements of the subspace, as in particular $\variety(\ideal(V))=V$. We say that an ideal $I$ has a \emph{basis} if $I = \langle S \rangle$ for some $S\subseteq \F_2[\vec X]$. By Hilbert's basis theorem, an ideal in $\F_2[\vec X]$ necessarily has a basis. Note that $\variety(S)= \variety(\langle S\rangle)$, and hence for a variety $V$ defined by a set of polynomials $S$ it suffices to represent $V$ as $\variety(\langle S\rangle)$. The union and intersection of varities corresponds to the product and sum of their ideals, respectively:
\[\variety(I) \cup \variety(J) = \variety (I\cdot J) \qquad\qquad \variety(I) \cap \variety(J) = \variety (I + J).\]
Moreover, given bases for $I$ and $J$ we can construct a basis for $I\cdot J$ and $I + J$, as
\[ \langle S \rangle \cdot \langle T \rangle = \langle \{ fg \mid f \in S, g \in T \} \rangle \qquad \qquad  \langle S \rangle + \langle T \rangle = \langle S \cup T \rangle.\]

We will be interested in a particular type of basis called a Gr\"{o}bner basis, defined with respect to a monomial order. Recall that a \emph{monomial order} $\succcurlyeq$ is a total well-order on monomials $m,n$ such that for any monomials $m,n,w$, $m\succcurlyeq n \iff mw\succcurlyeq nw$. A Gr\"{o}bner basis for a polynomial ideal $I$ over the monomial ordering $\succcurlyeq$ is a particular type of basis for $I$ over which reduction mod $I$ yields unique normal forms. For a particular monomial order, the \emph{reduced} Gr\"{o}bner basis of an ideal, obtained by mutually reducing each element of the Gr\"{o}bner basis with respect to one another, is unique. Methods for computing (reduced) Gr\"{o}bner bases exist \cite{buchberger,f4} and have been the subject of much study in symbolic computation --- though they require \emph{exponential} time unlike the Gaussian elimination used in ARA.

Given a polynomial ideal $I$ in disjoint variable sets $\vec X, \vec Y$, the \emph{elimination ideal} of $\vec X$ is  $I_{\vec Y} \defeq I \cap \F_2[\vec Y]$ which can be viewed as the polynomial relations on $\vec Y$ which are implied by $I$. Elimination ideals, which correspond to projection from the perspective of affine varieties, can be conveniently computed via Gr\"{o}bner basis methods. An elimination order for $\vec X$ is a monomial order such that any monomial involving a variable in $\vec X$ occurs earlier in the order than a monomial which does not contain a variable in $\vec X$. A standard result states that if $G$ is a Gr\"{o}bner basis for some ideal $I\subseteq \F_2[\vec X, \vec Y]$ in a monomial order eliminating $\vec X$, then $G\cap \F_2[\vec Y]$
is a Gr\"{o}bner basis for $I \cap \F_2[\vec Y]$.

As a final point we consider whether an ideal $I$ contains \emph{all} polynomial relations implied by the affine variety $\variety(I)$. In the case of algebraically closed fields this is the case, as $\ideal(\variety(I)) = I$ due to Hilbert's Nullstellensatz. While $\F_2$ is not algebraically closed and hence the Nullstellensatz fails, when considering \emph{multilinear} ideals --- ideals in $\F_2[\vec X]/\langle X_i^2 - X_i\mid X_i\in \vec X\rangle$ --- we can recover equality from Hilbert's \emph{strong} Nullstellensatz. Given an ideal $I$ in some polynomial ring, the \emph{radical} of $I$, denoted $\sqrt{I}$ is defined as the ideal 
\[
	\{f \mid f^m \in I \text{ for some integer } m\geq 1\}
\]
Hilbert's \emph{strong} Nullstellensatz asserts that $\ideal(\variety(I)) = \sqrt{I}$. As shown in \cite{gpc11}, over $\F_2$ an ideal's radical is obtained by adjoining the field equations $X_i^2 - X_i$, or equivalently reduce the ideal to \emph{multilinear} polynomials. %We give a proof below.
\begin{proposition}[\cite{gpc11}]\label{prop:radical}
	Let $I$ be an ideal over $\F_2[\vec X]$. Then \[\ideal(\variety(I)) = \sqrt{I} = I + \langle X_i^2 - X_i\mid X_i\in \vec X\rangle.\]
\end{proposition}

\Cref{prop:radical} implies in particular that our straightforward methods are in fact \emph{precise}, \emph{despite working over a non-algebraically closed field.}

\subsection{The $\mathbf{Pol}$ domain}

Given a unitary $U$ such that $U\ket{\vec x} = \ket{f_1(\vec x), f_2(\vec x), \dots, f_n(\vec x)}$ where $f_i$ are polynomial equations, the set of non-zero transitions $\bra{\vec x'}U\ket{\vec x} \neq 0$ of $U$ forms an affine variety over $\F_2^{2n}$ --- in particular, $\vec{x'}\in\supp{U\ket{\vec x}}$ if and only if $(\vec{x}',\vec {x})\in\variety(\{X_1' \oplus f_1(\vec X), \dots, X_n' \oplus f_n(\vec X)\})$. Given an affine variety $V$ in two vocabularies $\vec X', \vec X$ which contains all non-zero transitions of a quantum WHILE program, we may hence represent $V$ precisely as the set of polynomial relations $I=\ideal(V)$. We call this the $\mathbf{Pol}[\vec X'\sep \vec X]$ domain, whose elements are (Boolean) polynomial ideals $I$ over two (or more) vocabularies and whose concretization is given by $\variety(I)$. We implicitly adjoin the field equations $I_0=\langle X_i^2 - X_i,X_i'^2 - X'_i\mid X_i\in \vec X\rangle$ to elements of $\mathbf{Pol}$ ``under the hood'' as is standard in dealing with ideals over $\F_2$ \cite{bd09}.

Elements $I,J$ of $\mathbf{Pol}$ are stored as reduced G\"{o}bner bases, which allows checking equality of transition ideals. The domain $\mathbf{Pol}$ is naturally equipped with an order corresponding to the (reverse) subset relation on ideals $I \sqsubseteq J \iff I \supseteq J$, as well as greatest and least elements $\top = \emptyset$ and $\bot = \langle 1\rangle$,  respectively, where $\top$ is the \emph{least} precise element. The reverse order is chosen so that joins correspond to the union of varieties. In that regard we define $I \sqcup J = I\cdot J$ and $I\sqcap J = I + J$, where the concrete representation is again as a reduced Gr\"{o}bner basis. Projection $\exists \vec X.I[\vec X\sep \vec Y]$ corresponds to the elimination ideal $I_{\vec Y}$, which as discussed in the previous section can be implemented by computing a Gr\"{o}bner basis over an elimination order and projecting out $\vec X$. We can then define sequential composition as in affine subspaces as 
\[
	I[\vec X'\sep \vec X] \compose J[\vec X'\sep \vec X] = \exists \vec X''.I[\vec X''\sep \vec X] \sqcap J[\vec X'\sep \vec X''].
\]
The identity transition relation $1= \langle \vec X' \oplus \vec X\rangle$ once again serves as the neutral element with respect to composition.

Kleene iteration is surprisingly easy to define in our case, as we are working over the finite field $\F_2$ with polynomials which are in fact multilinear. Specifically, it can be observed that $\F_2[\vec X', \vec X]/I_0$ is finite, and so any ascending chain of ideals over this ring necessarily stabilizes. We can hence define the Kleene iteration $I^\kleene$ of an ideal similarity to the affine subspace domain as the limit of the sequence $\{I_i\}_{i=0}^\infty$ where
\[
	I_0 = \bot \qquad\qquad  I_{i+1} = I_i \sqcup (I_i \compose I)
\]
which again satisfies $I_{i}\sqsubseteq I_{i+1}$ for all $i\geq 0$. This avoids the need for solving recurrence relations as in most work on polynomial invariant generation \cite{rk04,ck24}, and is a significantly simplifying assumption which leads to easy implementation of non-linear reasoning in quantum programs.

\begin{remark}
	We use the ideal product for joins rather than the generally more precise ideal \emph{intersection} $I \cap J$, as the product is generally faster to compute \cite{cox}, and by \cref{prop:radical} and the fact that $\sqrt{I\cdot J} = \sqrt{I \cap J}$ asserts that the product is equal to the intersection in our case. We ran experiments with both and while we did not find a significant difference in computation time, they consistently produced the same ideal every time as expected.
\end{remark}

\subsection{A non-linear phase folding algorithm}

The $\mathbf{Pol}$ domain now suffices for a drop-in replacement of the affine $\mathbf{KS}$ domain in the phase folding algorithm of \cref{sec:fara}. We extend conditions of phases from linear functionals $f(\vec x)$ to \emph{polynomial} equations $f(\vec x)$, of which the linear functionals form a subset. Reduction of a phase condition modulo a transition ideal $I$ represented as a reduced Gr\"{o}bner basis $G$ amounts to multivariate polynomial division by $G$ which was previously noted to produce a unique canonical form $f'\in \F_2[\vec X'\sep \vec X]/I$. We denote the resulting analysis $\mathbf{PF}_{\mathrm{Pol}}\sem{\cdot}$.

\begin{theorem}
	Let $\mathbf{PF}_{\mathrm{Pol}}\sem{T} = (t, I)$. Then for every $(f, L)\in t$, the phase gates at locations $\ell\in L$ in $T$ can be safely merged into a single gate, or eliminated if $f=0$.
\end{theorem}

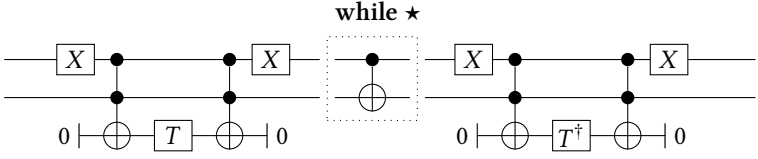
\begin{figure}
\begin{tikzpicture}
	\draw[] (-1,0.5) -- (9,0.5) node{};
	\node[] at (-0.2,0) {0};
	\node[] at (2.7,0) {0};
	\draw[] (0,0) node{|} -- (2.5,0) node{|};
	\draw[] (-1,1) -- (9,1) node{};
	\draw[fill=white] (-0.3,0.8) rectangle (0.2,1.2) node[pos=.5] {$X$};
	\node[circle,fill=black,inner sep=0pt,minimum size=5pt] (a) at (0.5,1) {};
	\node[circle,fill=black,inner sep=0pt,minimum size=5pt] at (0.5,0.5) {};
	\node[circle,draw=black,inner sep=0pt,minimum size=10pt] (b) at (0.5,0) {};
	\draw[] (a) -- (b.south) node {};
	\draw[fill=white] (1,-0.2) rectangle (1.5,0.2) node[pos=.5] {$T$};
	\node[circle,fill=black,inner sep=0pt,minimum size=5pt] (c) at (2,1) {};
	\node[circle,fill=black,inner sep=0pt,minimum size=5pt] at (2,0.5) {};
	\node[circle,draw=black,inner sep=0pt,minimum size=10pt] (d) at (2,0) {};
	\draw[] (c) -- (d.south) node {};
	\draw[fill=white] (2.3,0.8) rectangle (2.8,1.2) node[pos=.5] {$X$};
	\node[circle,fill=black,inner sep=0pt,minimum size=5pt] (x) at (3.9,1) {};
	\node[circle,draw=black,inner sep=0pt,minimum size=10pt] (y) at (3.9,0.5) {};
	\draw[] (x) -- (y.south) node {};
	\node[] at (5.1,0) {0};
	\node[] at (8,0) {0};
	\draw[] (5.3,0) node{|} -- (7.8,0) node{|};
	\draw[fill=white] (5,0.8) rectangle (5.5,1.2) node[pos=.5] {$X$};
	\node[circle,fill=black,inner sep=0pt,minimum size=5pt] (e) at (5.8,1) {};
	\node[circle,fill=black,inner sep=0pt,minimum size=5pt] at (5.8,0.5) {};
	\node[circle,draw=black,inner sep=0pt,minimum size=10pt] (f) at (5.8,0) {};
	\draw[] (e) -- (f.south) node {};
	\draw[fill=white] (6.3,-0.2) rectangle (6.8,0.2) node[pos=.5] {$T^\dagger$};
	\node[circle,fill=black,inner sep=0pt,minimum size=5pt] (g) at (7.3,1) {};
	\node[circle,fill=black,inner sep=0pt,minimum size=5pt] at (7.3,0.5) {};
	\node[circle,draw=black,inner sep=0pt,minimum size=10pt] (h) at (7.3,0) {};
	\draw[] (g) -- (h.south) node {};
	\draw[fill=white] (7.6,0.8) rectangle (8.1,1.2) node[pos=.5] {$X$};
	\draw[draw=white, line width = 2mm] (3.3,0.2) rectangle (4.5,1.3) node[above=0mm] {$\mathbf{while}\;  \star\qquad\quad$};
	\draw[dotted] (3.3,0.2) rectangle (4.5,1.3) node[pos=.5] {};
\end{tikzpicture}
\caption{A circuit with eliminable $\tgate$ gates. The loop satisfies the invariant $(1\oplus x_0')x_1' = (1\oplus x_0)x_1$.
}\label{fig:cool2}
\end{figure}

\begin{example}
\Cref{fig:cool2} gives an example of a looping quantum program which can be optimized by using a non-linear loop invariant. The loop satisfies the invariant $(1 \oplus x_0')x_1' = (1\oplus x_0)x_1$, as the second qubit is not modified if $x_0=0$. Noting that the $\tgate$ gates apply phases of $\omega^{(1\oplus x_0)x_1}$ and $(\overline{\omega})^{(1\oplus x_0')x_1'}$, respectively, the total phase contributed is $(\omega\overline{\omega})^{(1\oplus x_0)x_1}=1$ and both can hence be eliminated.

Our algorithm proceeds to compute the loop invariant by first constructing the transition ideal $\langle x_0' \oplus x_0, x_1' \oplus x_0 \oplus x_1\rangle$ of the loop body, corresponding to the $\cxgate:\ket{x_0,x_1}\mapsto \ket{x_0, x_1\oplus x_0}$ gate, and then taking the Kleene closure
\[\langle x_0' \oplus x_0, x_1' \oplus x_1 \oplus x_0\rangle^\kleene = \langle x_0' \oplus x_0, x_1' \oplus x_1\rangle\sqcup \langle x_0' \oplus x_0, x_1' \oplus x_1 \oplus x_0\rangle.\] Computing a Gr\"{o}bner basis of the right hand side gives $\langle x_0' \oplus x_0, x_1' \oplus x_1 \oplus x_0x_1' \oplus x_0x_1\rangle$, which notably reduces $(1\oplus x_0')x_1'$ to $(1 \oplus x_0)x_1$. After eliminating both $\tgate$ gates, the remaining gates can be canceled via basic gate cancellations.
\end{example}

\section{Increasing the precision of abstract transformers}\label{sec:rewrite}

It was previously noted that composition of transition relations (affine or otherwise) does \emph{not} produce the most precise relation for a control-flow path, due to the effects of \emph{interference} on classical states \emph{in superposition}. In this section, we show that precise transition relations of entire circuits can be generated automatically by symbolically identifying and reducing interfering paths using the \emph{sum-over-paths} technique. Such techniques are useful for generating abstract transformers for gates which are defined as circuits over some small set of basic gates.

\subsection{Symbolic path integrals}

A \emph{path integral} or \emph{sum} is a representation of a linear operator $\Psi:\C^{2^n}\rightarrow \C^{2^m}$ having the form
\begin{equation}
  \label{eq:pathsum1}
	\Psi:\ket{\vec{x}} \mapsto \sum_{\vec{y}\in\F_2^{k}}\Phi(\vec x, \vec y)\ket{f(\vec{x}, \vec{y})}
\end{equation}
where $\Phi : \F_2^{n+k}\rightarrow \C$ and $f:\F_2^{n+k}\rightarrow \F_2^n$ are the \emph{amplitude} and \emph{transition} functions, respectively. Intuitively, a path sum represents a linear operator (e.g. a quantum circuit) as a set of \emph{classical} transitions or \emph{paths} from $\ket{\vec x}$ to $\ket{f(\vec x, \vec y)}$ which are taken in superposition with amplitudes $\Phi(\vec x, \vec y)$. It was shown in \cite{a18} that by restricting to \emph{balanced} sums where $\Phi(\vec x, \vec y)=re^{i P(\vec{x}, \vec{y})}$ for some global normalization constant $r\in\C$, the path sum admits symbolic representation by multi-linear polynomials $P\in\R[\vec X, \vec Y]$ and $f = (f_1,\dots, f_n)$, $f_i\in\F_2[\vec X, \vec Y]$, which is moreover polynomial-time computable from a quantum circuit or channel. In particular, for gate sets which do not include \emph{non-linear classical} functions --- a class which includes, for instance, Clifford+$T$ and the exactly universal gate set $\{\cxgate, \rzgate(\theta),\rxgate(\theta) := \hgate\rzgate(\theta)\hgate\}$ --- the number of non-zero terms in $P$ and $f$ remains bounded polynomially in the size of the circuit.
\begin{example}
	Recall from \cref{sec:prelim} that the $\xgate$, $\tgate$, and $\hgate$ gates can be expressed as balanced path sums $\xgate: \ket{x} \mapsto \ket{1 \oplus x}$, $\tgate:\ket{x} \mapsto \omega \ket{x}$, and  $\hgate:\ket{x}\mapsto \frac{1}{\sqrt{2}}\sum_{y\in\F_2}(-1)^{xy}\ket{y}$.
	A path integral for the circuit $U=(\hgate\tgate\hgate)\otimes\xgate$ can be computed by composing the actions of each individual gate: 
	\begin{align*}
		U\ket{x_1,x_2} 
			= (\hgate\tgate\hgate\ket{x_1})\otimes (\xgate\ket{x_2}) 
			&=\left(\frac{1}{\sqrt{2}}\sum_{y_1}(-1)^{x_1y_1}(\hgate\tgate\ket{y_1})\right)\otimes \ket{1 \oplus x_2} \\
			&= \left(\frac{1}{\sqrt{2}}\sum_{y_1}(-1)^{x_1y_1}\omega^{y_1}(\hgate\ket{y_1})\right)\otimes \ket{1 \oplus x_2} \\
			&= \left(\frac{1}{2}\sum_{y_1,y_2}(-1)^{x_1y_1}\omega^{y_1}(-1)^{y_1y_2}\ket{y_2}\right)\otimes \ket{1 \oplus x_2} \\
			& = \frac{1}{2}\sum_{y_1,y_2}(-1)^{x_1y_1 + y_2y_1}\omega^{y_1}\ket{y_2,1 \oplus x_2}.
	\end{align*}
\end{example}
\noindent We use the notation $\ps{\Psi}$ to denote a path sum representation of $\Psi$ and $\ps{\Psi}\compose\ps{\Psi}'$ to denote the composition of path sums $\ps{\Psi}$, $\ps{\Psi'}$ with compatible dimensions as above. We also drop domains for variables in summations, as all variables are $\F_2$-valued.

A sound rewriting theory for balanced sums was further given in \cite{a18}. The rewrite system of \cite{a18} is presented in \cref{fig:cliffrewrite}, where rewrite rules are applied to the right-hand side of an expression $\ket{\vec{x}} \mapsto \sum_{\vec{y}}e^{i P(\vec{x}, \vec{y})}\ket{f(\vec{x}, \vec{y})}$. rewriting in the sum-over-paths can be interpreted as \emph{contracting interfering paths}. For instance, the \ref{eq:i} rule of \cref{fig:cliffrewrite} arises from the fact that in the path integral
\[
	\sum_{\vec x,y,z}(-1)^{y\left(z \oplus P(\vec x)\right)}e^{iQ(\vec x,z)}\ket{f(\vec x,z)} = \sum_{\vec x,z}\left(\sum_{y}(-1)^{y\left(z \oplus P(\vec x)\right)}\right)e^{iQ(\vec x,z)}\ket{f(\vec x,z)},
\]
the $y=0$ and $y=1$ paths have opposite phase and hence cancel whenever $z \oplus P(\vec x) = 0$. Hence we can safely assume $z=P(\vec x)$ in any paths with non-zero amplitude. We say that this rule is a \emph{contraction} on $y$ and witnesses the constraint $z=P(\vec x)$. More generally, given a sum of the form
\[
	\sum_{\vec x,y}(-1)^{yP(\vec x)}e^{iQ(\vec x)}\ket{f(\vec x)}=\sum_{\vec x}\left(\sum_{y}(-1)^{yP(\vec x)}\right)e^{iQ(\vec x)}\ket{f(\vec x)},
\]
we see that any path where $P(\vec x) \neq 0$ has zero amplitude, as the two paths corresponding to $y=0$ and $y=1$ again have opposite phase and destructively interfere. We again say that contraction on $y$ witnesses the constraint $P(\vec x)=0$, though we can not in general rewrite the integral as $P(\vec x) = 0$ may not be solvable by substitution. The \ref{eq:i} rule arises as a particular instance of this binary interference scheme where $P(\vec x) = 0$ admits a solution of the form $z = P'(\vec x)$ where $z$ is bound by a summation. 

We say an instance of \ref{eq:i} is \emph{affine} if $P$ is some affine expression in its free variables --- for instance, $P(\vec x)=1\oplus x_1\oplus x_2\oplus x_5$ --- and more generally that it has degree $k$ if $\deg (P) = k$. While the rewriting system of \cref{fig:cliffrewrite} is not complete nor even confluent for circuits over Clifford+$T$, it is complete for Clifford circuits even when restricting \ref{eq:i} to affine instances \cite{v21}.

\begin{figure}
\begin{align}
	\sum_{\vec x,y}e^{iQ(\vec x)}\ket{f(\vec x)} &\longrightarrow_{\mathrm{Cliff}} 2\sum_{\vec x}e^{iQ(\vec x)}\ket{f(\vec x)} \tag*{(E)}\label{eq:e} \\
	\sum_{\vec x,y,z}(-1)^{y\left(z \oplus P(\vec x)\right)}e^{iQ(\vec x,z)}\ket{f(\vec x,z)} &\longrightarrow_{\mathrm{Cliff}} 2\sum_{\vec x}e^{iQ(\vec x,\overline{P}(\vec x)}\ket{f(\vec x,P(\vec x))} \tag*{(H)}\label{eq:i} \\
	\sum_{\vec x,y}i^y(-1)^{yP(\vec x)}e^{iQ(\vec x)}\ket{f(\vec x)} &\longrightarrow_{\mathrm{Cliff}} \omega\sqrt{2}\sum_{\vec x}(-i)^{\overline{P}(\vec x)}e^{iQ(\vec x)}\ket{f(\vec x)} \tag*{($\omega$)}\label{eq:u}
\end{align}
\caption{The rewrite rules of \cite{a18}. In all rules above $P$ is a Boolean polynomial and $z,y\notin FV(P)$ and $\overline{P}$ lifts $P$ to an equivalent polynomial over $\R$ via the sound translation $\overline{P \oplus Q} = \overline{P} + \overline{Q} - 2\overline{P}\overline{Q}$.}
\label{fig:cliffrewrite}
\end{figure}

\subsection{Generating precise transition relations with the sum-over-paths}

We now describe how path sums can be used to generate sound abstractions of the classical semantics $\powerset{\mathcal{C}_\mathcal{Q}} \rightarrow \powerset{\mathcal{C}_\mathcal{Q}}$ of a circuit, and how the rewrite rules of \cref{fig:cliffrewrite} can be used to increase the precision of the abstraction.

As noted above, the sum-over-paths encodes a linear operator as a set of classical transitions 
\[
	\ket{\vec x} \mapsto \{ \ket{f(\vec x, \vec y)} \mid \vec y \in \F_2^k\}.
\]
for some vector $f=(f_1,\dots, f_n)$ of polynomials $f_i$. In particular, given a path sum expression $\Psi:\ket{\vec{x}} \mapsto \sum_{\vec{y}\in\F_2^{k}}\Phi(\vec x, \vec y)\ket{f(\vec{x}, \vec{y})}$, there exists $\vec y\in F_2^k$ such that the relation $\vec x ' = f(\vec x, \vec y)$ holds whenever $\bra{\vec x'} \Psi \ket{\vec x} \neq 0$. Hence, a sound abstraction of $\Psi$ in the $\mathbf{Pol}$ domain is the polynomial ideal \[\exists \vec Y. \langle \vec X' \oplus f(\vec X, \vec Y)\rangle=\exists \vec Y. \langle X_1' \oplus f_1(\vec X, \vec Y),\dots, X_n' \oplus f_n(\vec X, \vec Y) \rangle.\]

\begin{definition}
	Given a path sum $\ps{\Psi}=\ket{\vec{x}} \mapsto \sum_{\vec{y}\in\F_2^{k}}\Phi(\vec x, \vec y)\ket{f(\vec{x}, \vec{y})}$, the abstraction of $\ps{\Psi}$ in the $\mathbf{Pol}$ domain is
	\[
		\alpha\left(\ps{\Psi}\right) \defeq \exists \vec Y. \langle \vec X' \oplus f(\vec X, \vec Y)\rangle 
	\]
\end{definition}
\begin{proposition}\label{prop:pssound}
	For any path sum representation $\ps{\Psi}$, $\alpha\left(\ps{\Psi}\right)$ is a sound abstraction of $\Psi$.
\end{proposition}
\begin{proposition}\label{prop:compose}
Let $\ps{\Psi}$ and $\ps{\Psi'}$ be (composable) path sums. Then \[\alpha(\ps{\Psi}\compose\ps{\Psi}') \sqsubseteq \alpha(\ps{\Psi})\compose\alpha(\ps{\Psi'}).\]
\end{proposition}

Note that in the case when $\deg(f_i) \leq 1$ for all $i$, then the polynomial mapping is in fact an affine map $(A, \vec c)$, and can be likewise soundly abstracted in a domain of affine relations as $\vec X' \oplus  A(\vec X, \vec Y) \oplus \vec c$. The affine transition relations for gates in \cref{fig:ara} arise from the specifications of each gate as a path sum in this way. If $\deg f_i > 1$, we can always soundly abstract the transition $X_i' = f_i(\vec X, \vec Y)$ in an affine relational domain simply as $\top_i$.

Given a path sum $\ps{\Psi}$, if $\ps{\Psi}\longrightarrow_{\mathrm{Cliff}} \ps{\Psi}'$, then by the soundness of $\longrightarrow_{\mathrm{Cliff}}$ both represent the same linear operator $\Psi$. In particular, $\alpha(\ps{\Psi}')$ gives a sound abstraction of $\Psi$, which may be \emph{more precise} than $\alpha(\ps{\Psi})$ as the next example shows.

\begin{example}
	Recall that the $\hgate$ is self-inverse, and in particular $\hgate\hgate=\igate$, which leads to imprecision in the phase folding analysis since $\abssem{\hgate\hgate} = \abssem{\hgate}\compose\abssem{\hgate} = \top$. If we compute the sum-over-paths representation of $\hgate\hgate$ by composition, we see that
\[
	\ps{\hgate}\compose\ps{\hgate}=\ket{x}\mapsto\frac{1}{2}\sum_{y,z}(-1)^{xy + yz}\ket{z}.
\]
Noting that $\sum_{y,z}(-1)^{xy + yz}\ket{z}=\sum_{y,z}(-1)^{y(z\oplus x)}\ket{z} \longrightarrow_{\mathrm{Cliff}} 2\ket{x}$ by \ref{eq:i}, rewriting the above expression gives the precise classical semantics $\hgate\hgate:\ket{x}\mapsto\ket{x}$.
\end{example}

We show next that rewriting a path sum can only \emph{increase} the precision of the abstraction, as it results from \emph{eliminating} some interfering paths:

\begin{proposition}\label{prop:precise}
Let $\ps{\Psi}$ by a symbolic path integral and suppose $\ps{\Psi}\longrightarrow_{\mathrm{Cliff}} \ps{\Psi}'$. Then \[\alpha(\ps{\Psi}')\sqsubseteq \alpha(\ps{\Psi}).\]
\end{proposition}
\begin{proof}
For the \ref{eq:e} and \ref{eq:u} rules the proof is trivial as the output expression is unchanged in either case. For the \ref{eq:i} rule it can be observed that \begin{align*}\alpha\left(\ket{\vec x}\mapsto \sum_{\vec y,z,z'}(-1)^{z'\left(z \oplus P(\vec x, \vec y)\right)}e^{iQ(\vec x,\vec y,z)}\ket{f(\vec x, \vec y,z)}\right) &= \exists \{\vec Y,Z,Z'\}. \langle \vec X' \oplus f(\vec X, \vec Y, Z)\rangle \\ &\sqsupseteq \exists \vec Y.\langle \vec X' \oplus f(\vec X,\vec Y, P(\vec X))\rangle\end{align*}
\end{proof}

Given a circuit or control-flow path $\pi$ in a quantum WHILE program, by \cref{prop:pssound,prop:compose,prop:precise} we can synthesize an abstraction of $\pi$ which is \emph{at least as precise} as (and often more precise than) the methods of \cref{sec:ara,sec:pra} by computing a path sum representation $\ps{\pi}$ of $\pi$ through standard techniques (e.g. \cite{a18}), reducing it using \cref{fig:cliffrewrite}, then abstracting it into a transition formula using a relevant domain (e.g. $\mathbf{Pol}$). We summarize this process in \cref{alg:two} below:
\begin{algorithm}
\begin{algorithmic}
\Function{Synthesize-transformer}{$\pi$}
	\State Compute a path sum $\ps{\pi}$ expression of $\pi$
	\State \textbf{while} $\ps{\pi} \longrightarrow_{\mathrm{Cliff}} \ps{\pi}'$, set $\ps{\pi} := \ps{\pi}'$
	\State \Return $\alpha(\ps{\pi})$
\EndFunction
\end{algorithmic}
\caption{Synthesis of abstract transformers.}\label{alg:two}
\end{algorithm}

\Cref{alg:two} can be used to generate (relatively) precise relational abstractions of basic blocks in a quantum WHILE program. As the infinite sum of path integrals however is not bounded, a suitable, terminating representation of an entire program in terms of path sums is not easily possible.

\paragraph{Complexity}
While the \emph{computation} of the path integral along a control-flow path is polynomial time in the number of qubits, \emph{reduction of the path integral may in fact take exponential time} as the degree of the constituent polynomials may increase during reduction. If however rewriting is restricted to \emph{affine} relations, \cref{alg:two} terminates in time $O(m^{c}|\pi|)$ where $|\pi|$ is the length of the path, $m\leq n + |\pi|$ is the number of variables in the path sum, and $c$ is the maximum degree of polynomials appearing in the path sum. For Clifford+$T$ circuits, $c=3$ \cite{a18}.

\subsection{Phase folding modulo path integral rewriting}

\Cref{alg:two} can be used in phase folding to generate effective summaries of the classical semantics for basic blocks. However, as in \cref{sec:ara} we also wish to know which phase gates \emph{inside} the block can be merged.

\Cref{alg:one} presents the \textsc{Strengthen} algorithm, which gathers the constraints witnessed through rewriting of the block's path integral, and then uses them to further reduce the phase conditions within the block. \textsc{Strengthen} attempts to construct as large of a transition ideal as possible over the pre-, post-, and intermediate-variables in the block by repeatedly adding constraints $P(\vec x) = 0$ witnessed by the hypothetical contraction of $\sum_{\vec x,y}(-1)^{yP(\vec x)}e^{iQ(\vec x)}\ket{f(\vec x)}$ on $y$, before picking a specific contraction to apply (i.e. applying a rewrite rule).

\begin{algorithm}
\begin{algorithmic}
\Function{Strengthen}{$\pi, (t,A) := \mathbf{PF}\sem{\pi}$}
	\State Compute a path sum $\ps{\pi}$ expression of $\pi$
	\While{$\ps{\pi} \longrightarrow_{\mathrm{Cliff}} \ps{\pi}'$}
		\State Set $A := A \sqcap \langle P(\vec X) = 0\rangle$ whenever $\ps{\pi}$ has the form $\sum_{\vec x,y}(-1)^{yP(\vec x)}e^{iQ(\vec x)}\ket{f(\vec x)}$
		\State $\ps{\pi} := \ps{\pi}'$
	\EndWhile
	\State $t := \uplus_{(f,S)\in t} \left(\mathrm{reduce}(A, f), S\right)$
	\State \Return $(t, A)$
\EndFunction
\end{algorithmic}
\caption{Strengthening phase folding by symbolic interference analysis.}\label{alg:one}
\end{algorithm}

\begin{example}\label{ex:badnew}
	Consider the circuit $T:=\tgate^\dagger\hgate\hgate\tgate.$ Denoting by $x,x'$ the pre- and post-states, and $y$ the output of the first Hadamard gate, the phase folding analysis with either domain (affine or polynomial) returns
\[
	\mathbf{PF}\sem{T} = (t, A) = \left([x \mapsto \{\ell\}, x' \mapsto \{\ell'\}], \top\right).
\]
where $x \mapsto \{\ell\}$ denotes the pair $(x, \{\ell\})$.
In particular, the analysis shows that the $\tgate$ gate at location $\ell$ rotates the phase if $x=1$, and the $\tgate^\dagger$ at location $\ell'$ rotates the phase if $x'=1$, and no relation between the pre- and post-state holds. Hence, neither gate can be merged with the other, despite the circuit being equal to the identity. Computing the path sum representation of $T$ however, we see that
\[
	T : \ket{x} \mapsto \frac{1}{2}\sum_{y,x'}(-1)^{xy + yx'}\omega^{x - x'}\ket{x'}.
\]
The sum can be contracted on $y$, which witnesses the equality $x'=x$:
\[
	\frac{1}{2}\sum_{y,x'}(-1)^{xy + yx'}\omega^{x - x'}\ket{x' } = \sum_{y,x'}(-1)^{y(x \oplus x')}\omega^{x - x'}\ket{x' } \xrightarrow{\;\textrm{\ref{eq:i}}\;}_{\mathrm{Cliff}} \omega^{x - x}\ket{x} = \ket{x}.
\]
The \textsc{Strengthen} algorithm uses this equality to strengthen the transition ideal by setting $A:=\top \sqcap \langle x' \oplus x\rangle = \langle x' \oplus x\rangle$. Finally, \textsc{Strengthen} reduces all phase conditions modulo $A$, where in particular $\mathrm{reduce}(A, x) = x = \mathrm{reduce}(A, x')$ and the algorithm hence returns the stronger result $\left([x \mapsto \{\ell, \ell'\}], \langle x = x'\rangle \right)$ which shows the $\tgate$ and $\tgate^\dagger$ gate can be merged.
\end{example}

In the above example, the optimization could be achieved in a forward-manner similar to \cref{sec:fara}. In particular, after the second Hadamard gate the path integral is 
\[
	\ket{x} \mapsto \frac{1}{2}\sum_{y,x'}(-1)^{xy + yx'}\omega^{x}\ket{x'} = \ket{x}
\]
which reduces via \ref{eq:i} to $\ket{x}$, at which point applying the final $\tgate^\dagger$ to $\ket{x}$ will merge with the first $\tgate$. \textsc{Strengthen} works instead by first computing a path integral for the \emph{entire block}, and then applying path sum reductions to strengthen the analysis. This is for technical reasons owing to the fact that the rewriting system of \cref{fig:cliffrewrite} is non-confluent in general. The example below illustrates this fact and its impact on phase folding.

\begin{example}\label{ex:bad}
Consider the circuit below, which occurs in the $8$-bit adder circuit of \cite{ttk10}. 
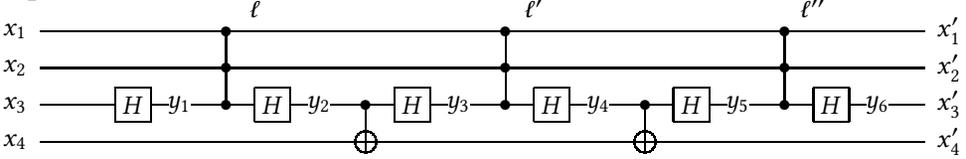
\[
\Qcircuit @C=.4em @R=.02em @! {
\lstick{x_1} & \qw & \qw & \qw & \ctrl{2} & \ustick{\small \!\!\!\!\!\!\!\!\ell}\qw & \qw & \qw & \qw & \qw & \ctrl{2} & \ustick{\small \!\!\!\!\!\!\!\!\ell'}\qw & \qw  & \qw & \qw & \qw & \ctrl{2} & \ustick{\small \!\!\!\!\!\!\!\!\ell''}\qw & \qw & \rstick{x_1'}\qw \\
\lstick{x_2} & \qw & \qw & \qw & \ctrl{1} & \qw & \qw  & \qw & \qw & \qw & \ctrl{1} & \qw & \qw & \qw & \qw & \qw & \ctrl{1} & \qw & \qw & \rstick{x_2'}\qw \\
\lstick{x_3} & \qw & \gate{H} &\push{y_1}\qw & \ctrl{0} & \gate{H} &\push{y_2}\qw & \ctrl{1} & \gate{H} &\push{y_3}\qw & \ctrl{0} & \gate{H} &\push{y_4}\qw & \ctrl{1} & \gate{H} &\push{y_5}\qw & \ctrl{0} & \gate{H} &\push{y_6}\qw & \rstick{x_3'}\qw \\
\lstick{x_4} & \qw & \qw & \qw & \qw & \qw  & \qw & \targ & \qw & \qw  & \qw & \qw  & \qw & \targ & \qw & \qw & \qw & \qw & \qw & \rstick{x_4'}\qw
}
\]
Intermediate variables corresponding to the outputs of Hadamard gates are labeled in the circuit for convenience.
Recall that the $\cczgate$ gates depicted as \raisebox{0.6em}{$\Qcircuit @C=.4em @R=.3em @! { & \ctrl{2} & \qw \\ &  \ctrl{1} & \qw \\ &  \ctrl{0} & \qw}$} implement the diagonal transformation $\cczgate:\ket{x_1,x_2,x_3} \mapsto (-1)^{x_1x_2x_3}\ket{x_1,x_2,x_3}$. The result of the phase folding analysis, extended to this gate set, is a pair $(t, A)$ where
\begin{align*}
	t &= [x_1x_2y_1 \mapsto \{\ell\}, x_1x_2y_3 \mapsto \{\ell'\}, x_1x_2y_5 \mapsto \{\ell''\}], \\ A &= \langle x_1' = x_1, x_2'=x_2, x_3'=y_6,x_4'=x_4\oplus y_2\oplus y_4\rangle.
\end{align*}
Computing the circuit's path integral, we similarly get
\begin{align*}
	\ket{x_1,x_2,x_3,x_4}&\mapsto\sum_{\vec y\in \F_2^6}(-1)^{P(\vec x, \vec y)}\ket{x_1,x_2,y_6,x_4\oplus y_2 \oplus y_4}, \\
	P(\vec x, \vec y) &= y_1(x_3 \oplus x_1x_2 \oplus y_2) + y_3(y_2 \oplus x_1x_2 \oplus y_4) + y_5(y_4 \oplus x_1x_2 \oplus y_6)
\end{align*}
where the phase polynomial $P(\vec x, \vec y)$ is factored to show the three possible contractions on $y_1$, $y_3$, and $y_5$, witnessing the constraints $x_3 \oplus x_1x_2 \oplus y_2 = 0$, $y_2 \oplus x_1x_2 \oplus y_4=0$, and $y_4 \oplus x_1x_2 \oplus y_6=0$, respectively. \textsc{Strengthen} hence computes
\[
	A':= A \sqcap \langle x_3 \oplus x_1x_2 \oplus y_2, y_2 \oplus x_1x_2 \oplus y_4, y_4 \oplus x_1x_2 \oplus y_6\rangle.
\]
Applying the contraction on $y_1$ and substituting $y_2 \gets x_1x_2 \oplus x_3$ then yields
\[
	\sum(-1)^{y_3(x_3 \oplus y_4) + y_5(y_4 \oplus x_1x_2 \oplus y_6)}\ket{x_1,x_2,y_6,x_4\oplus x_3 \oplus x_1x_2 \oplus y_4}.
\]
Contraction on either $y_3$ or $y_5$ yields no new equations, as both $x_3 \oplus y_4\in A'$ and $y_4 \oplus x_1x_2 \oplus y_6\in A'$. Picking $y_3$ to contract on next and substituting $y_4\gets x_3$ yields
\[
	\sum(-1)^{y_5(x_3 \oplus x_1x_2 \oplus y_6)}\ket{x_1,x_2,y_6,x_4\oplus x_1x_2}.
\]
Again, $x_3 \oplus x_1x_2 \oplus y_6\in A'$ and the final contraction on $y_5$ produces the precise transition relation $\ket{x_1,x_2,x_3,x_4}\mapsto\ket{x_1,x_2,x_3\oplus x_1x_2,x_4\oplus x_1x_2}$. The final step is to reduce $t$ with respect to $A'$, which we can do by computing a reduced Gr\"{o}bner basis (in grevlex order)
\[
	A' = \langle x_3 \oplus y_2 \oplus x_1x_2, x_3 \oplus y_2 \oplus x_2x_3 \oplus x_2y_2, x_3 \oplus y_2 \oplus x_1x_3 \oplus x_1y_2, x_3\oplus y_4, y_2 \oplus y_6\rangle.
\]
Reducing $t$ modulo $A'$ then gives $[x_3y_1 \oplus y_1y_2 \mapsto \{\ell\}, x_3y_3 \oplus y_2y_3 \mapsto \{\ell'\}, x_3y_5 \oplus y_2y_5 \mapsto \{\ell''\}]$
and hence no gates can be merged.

If however the first contraction chosen was n $y_3$, we get the following path integral
\[
	\sum(-1)^{y_1x_3 + y_4(y_1 \oplus y_5) + y_5x_1x_2 + y_5y_6}\ket{x_1,x_2,y_6,x_4\oplus x_1x_2}.
\]
We now see that interference on $y_4$ witnesses the constraint $y_1 \oplus y_5=0$. Setting $A'' := A' \sqcap \langle y_1 \oplus y_5\rangle$ and reducing $t$ modulo $A''$ we get
\[
	[x_3y_3 \oplus y_2y_3 \mapsto \{\ell'\}, x_3y_1 \oplus y_1y_2 \mapsto \{\ell,\ell''\}]
\]
which in particular allows the elimination of the $\cczgate$ gates at locations $\ell$ and $\ell''$.
\end{example}

It remains a direction for future work to determine methods of both confluently rewriting the path sum, as well as to efficiently find all polynomial constraints witnessed by binary interference of the form $\sum_y(-1)^{yP(\vec x)}$.

\paragraph{Implementation}
In practice, \cref{alg:one} generates ideals which are too large for our implementation to feasibly compute reduced Gr\"{o}bner bases, as the number of intermediate variables is proportional to the length of the circuit block. Instead, we non-canonically reduce phase conditions with respect to the full set of generated equations, and hence our implementation theoretically misses some gates which can be merged with \cref{alg:one}. For small programs in which Gr\"{o}bner bases could be computed, we saw no difference in optimization between the two methods. To speed up reduction we also experimented with methods of \emph{linearizing} the ideal and reducing via Gaussian elimination, but saw little gain in performance.

\section{Experimental evaluation}

We implemented our optimizations in the open-source package \textsc{Feynman}\footnote{\href{https://github.com/meamy/feynman}{github.com/meamy/feynman}}. The sources and benchmarks used in the experiments are provided in the associated software artifact \cite{artifact}. We denote the affine optimization of \cref{sec:ara} by $\mathbf{PF}_{\mathrm{Aff}}$, and the polynomial optimization of \cref{sec:pra} combined with the \textsc{Strengthen} algorithm of \cref{sec:rewrite} by $\mathbf{PF}_{\mathrm{Pol}}$. We also ran experiments where \textsc{Strengthen} was limited to \emph{quadratic} relations, which we denote by $\mathbf{PF}_{\mathrm{Quad}}$, in order to ascertain the effectiveness of highly non-linear relations over those which characterize the Toffoli gate. In both the unbounded and quadratic cases, reduction in \textsc{Strengthen} was performed using non-canonical multivariate division, as computing a full Gr\"{o}bner basis caused almost all benchmarks to time out.

We performed two sets of experiments: \emph{program} optimization experiments which involve classical control, and straightline \emph{circuit} optimization experiments. All experiments were run in Ubuntu 22.04 running on an AMD Ryzen 5 5600G 3.9GHz processor and 64GB of RAM. Optimizations were given a $2$-hour TIMEOUT and $32$GB memory limit.

\subsection{Program optimization benchmarks}

For the program optimization benchmarks, a front-end for the quantum programming language openQASM 3 \cite{qasm} was written and used to perform optimizations on quantum programs which combine circuits and classical control. A subset of openQASM 3 which restricts loops to a single entry and exit point is modeled as a non-deterministic quantum WHILE program by (1) eliminating any classical computation, (2) replacing branch and loop conditions with $\star$, and (3) inlining \emph{gate} calls. Note that procedure calls are \emph{not} inlined whereas gate calls are, as most optimizations in practice occur across sub-circuit boundaries.

The results for our program optimization benchmarks are present in \cref{tab:programs}. As we could not find a set of suitable optimization benchmarks in openQASM 3, we tested our relational optimization on a selection of hand-written micro-benchmarks designed to test the ability to compute certain invariants and to perform the associated optimizations. The benchmarks are provided in the \textsc{Feynman} repository\footnote{\href{https://github.com/meamy/feynman/tree/ara/benchmarks/qasm3}{github.com/meamy/feynman/tree/ara/benchmarks/qasm3}}, and include the examples given throughout the paper -- notably \cref{fig:ruswhole} (RUS), \cref{fig:cool} (Loop-swap), and \cref{fig:cool2} (Loop-nonlinear). For the Grover benchmark, an instance of Grover's search on a $64$-bit function was generated and the loop ($\approx 10^9$ iterations) was modeled as a non-deterministic loop for optimization. The results in \cref{tab:programs} confirm that our methods are able to find some non-trivial optimizations in simple hybrid programs which have interesting classical control. Moreover, the experiments demonstrate that phase folding can be integrated into compilers targeting hybrid quantum/classical programming languages.

\begin{table}
\scriptsize
\begin{tabular}{lrrrrrrc} \toprule
Benchmark & $n$ & \multicolumn{1}{c}{Original} & \multicolumn{2}{c}{$\mathbf{PF}_{\mathrm{Aff}}$} & \multicolumn{2}{c}{$\mathbf{PF}_{\mathrm{Pol}}$} & Loop invariant \\ \hline
 & & \# $\tgate$  & $\tgate$ count & time (s) & \# $\tgate$ & time (s) \\
 \cmidrule(lr){4-5} \cmidrule(lr){6-7}
RUS & 3 & 16  & 10 & 0.30 & 8 & 0.35 & $\langle z' \oplus z\rangle$  \\
Grover & 129 & $1736\times 10^9$ & $1472\times 10^9$ & 1.98 & \multicolumn{2}{r}{TIMEOUT} & -- \\% \\
Reset-simple & 2 & 2 & 1 & 0.15 & 1 & 0.23 & -- \\%& Yes \\
If-simple & 2 & 2  & 0 & 0.18 & 0 & 0.16 & -- \\
Loop-simple & 2 & 2  & 0 & 0.17 & 0 & 0.16 & $\langle x' \oplus x, y \oplus y' \oplus xy \oplus xy' \rangle$  \\
Loop-h & 2 & 2  & 0 & 0.16 & 0 & 0.16 & $\langle y' \oplus y\rangle$ \\
Loop-nested & 2 & 3 & 2 & 0.17 & 2 & 0.18 & $\langle x' \oplus x\rangle$, $\langle x' \oplus x\rangle$  \\
Loop-swap & 2 & 2 & 0 & 0.30 & 0 & 0.20 & $\langle x' \oplus y' \oplus x \oplus y, x' \oplus xy \oplus xx' \oplus yx' \rangle$  \\
Loop-nonlinear & 3 & 30 & 18 & 0.44 & 0 & 0.26 & $\langle x' \oplus x, z' \oplus z, y' \oplus y \oplus xy \oplus xy'\rangle$ \\
Loop-null & 2 & 4 & 2 & 0.18 & 2 & 0.17 & $\langle x' \oplus x, y' \oplus y\rangle$ \\ \bottomrule
\end{tabular}
\vspace{2em}
\caption{Optimization of $\tgate$-count in hybrid openQASM 3.0 micro-benchmarks. All other non-diagonal gate counts are left unchanged. The loop invariant computed by $\mathbf{PF}_{\mathrm{Pol}}$ is also given.}\label{tab:programs}
\vspace*{-3em}
\end{table}

\subsection{Circuit optimization benchmarks}

\begin{table}
\scriptsize
\begin{tabular}{lrrrrrrrrrrrr} \toprule
Benchmark & $n$ & Original & \multicolumn{2}{c}{PyZX} & \multicolumn{2}{c}{$\mathbf{PF}_{\mathrm{Aff}}$} & \multicolumn{2}{c}{$\mathbf{PF}_{\mathrm{Quad}}$}& \multicolumn{2}{c}{$\mathbf{PF}_{\mathrm{Pol}}$} & \multicolumn{2}{c}{+FastTODD} \\ \midrule
 & & \# $\tgate$  &  \# $\tgate$ & time (s) &  \# $\tgate$ & time (s) &  \# $\tgate$ & time (s)&  \# $\tgate$ & time (s) & PyZX & $\mathbf{PF}_{\mathrm{Pol}}$  \\
\cmidrule(lr){3-3} \cmidrule(lr){4-5} \cmidrule(lr){6-7} \cmidrule(lr){8-9} \cmidrule(lr){10-11}\cmidrule(lr){12-13}
Grover$\_5$ &9&336&166&0.26& \textbf{148} &0.18& \textbf{148} &0.2&\textbf{0}&0.47&143&\textbf{0}\\
Mod $5\_4$ &5&28&8&0.01&8& <0.01 &8& <0.01&8&0.47&7&7\\
VBE-Adder$\_3$ &10&70&24&0.02&24&0.01&24&0.01&24&0.01&19&19\\
CSLA-MUX$\_3$ &15&70&62&0.03& \textbf{60} &0.01& \textbf{60} &0.01&\textbf{60}&0.02&39&39\\
CSUM-MUX$\_9$ &30&196&84&0.08&84&0.02&84&0.02&84&0.15&71&71\\
QCLA-Com$\_7$ &24&203&95&0.12& \textbf{94} &0.04& \textbf{94} &0.04&\textbf{94}&0.5&\textbf{59}&61\\
QCLA-Mod$\_7$ &26&413&237&0.38&237&0.17&237&0.19&237&39.13&\textbf{159}&161\\
QCLA-Adder$\_10$ &36&238&162&0.11&162&0.05&162&0.06&162&0.85&109&109\\
Adder$\_{8}$ &24&399&173&0.51&173&0.18&173&0.21&173&2.31&\textbf{119}&121\\
RC-Adder$\_{6}$ &14&77&47&0.04&47&0.02&47&0.02&47&0.05&37&37\\
Mod-Red$\_{21}$ &11&119&73&0.06&73&0.02&73&0.02&73&0.04&51&51\\
Mod-Mult$\_{55}$ &9&49&35&0.02&35&0.01&35&0.01&35&0.01&17&17\\
Mod-Adder$\_{1024}$ &28&1995&1011&3&1011&14.01& \textbf{1005} &14.16&\textbf{923}&36.34&--&--\\
GF($2^4$)-Mult &12&112&68&0.06&68&0.02&68&0.02&68&0.02&49&49\\
GF($2^5$)-Mult &15&175&115&0.08&115&0.03&115&0.04&115&0.04&81&\textbf{75}\\
GF($2^6$)-Mult &18&252&150&0.13&150&0.07&150&0.07&150&0.04&113&\textbf{111}\\
GF($2^7$)-Mult &21&343&217&0.25&217&0.11&217&0.11&217&0.09&155&\textbf{141}\\
GF($2^8$)-Mult &27&448&264&0.63&264&0.4&264&0.4&264&0.46&205&\textbf{203}\\
GF($2^9$)-Mult &24&567&351&0.5&351&0.37&351&0.38&351&0.44&257&\textbf{255}\\
GF($2^{10}$)-Mult &27&700&410&1.06&410&0.61&410&0.6&410&0.7&315&\textbf{313}\\
GF($2^{16}$)-Mult &48&1792&1040&5.5&1040&17.05&1040&16.87&1040&18.28&\textbf{797}&803\\
GF($2^{32}$)-Mult &96&7168&4128& 9.33m &4128& 67m &4128& 60m&4128&86m&--&--\\
Ham$\_{15}$ (low) &17&161&97&0.46&97&0.31&97&0.3&97&0.44&77&77\\
Ham$\_{15}$ (med) &17&574&212&1.8&212&0.3&212&0.31&\textbf{210}&1.59&\textbf{137}&140\\
Ham$\_{15}$ (high) &20&2457&1019&35.09&1019&21.33& \textbf{997} &21.23&\textbf{985}&2m&--&--\\
HWB$\_6$ &7&105&75&0.07&75&0.02&75&0.03&75&0.12&51&51\\
QFT$\_4$ &5&69&67&0.02&67&0.01& \textbf{65} &0.01&\textbf{65}&0.03&\textbf{53}&54\\
$\Lambda_3(X)$ &5&21&15&0.01&15& <0.01 &15&<0.01&15&<0.01&13&13\\
$\Lambda_4(X)$ &7&35&23&0.01&23& <0.01 &23& <0.01&23&<0.01&19&19\\
$\Lambda_5(X)$ &9&49&31&0.01&31& <0.01 &31&0.01&31&0.01&25&25\\
$\Lambda_{10}(X)$ &19&119&71&0.04&71&0.01&71&0.02&71&0.01&55&55\\
$\Lambda_3(X)$ (dirty) &5&28&16&0.01&16&0.01&16&0.01&16&0.01&13&13\\
$\Lambda_4(X)$ (dirty) &7&56&28&0.02&28&0.01& \textbf{24} &0.01&\textbf{24}&0.01&23&\textbf{20}\\
$\Lambda_5(X)$ (dirty) &9&84&40&0.03&40&0.01& \textbf{32} &0.01&\textbf{32}&0.01&33&\textbf{28}\\
$\Lambda_{10}(X)$ (dirty) &19&224&100&0.11&100&0.05& \textbf{72} &0.03&\textbf{72}&0.06&83&\textbf{68}\\ \midrule 
FP-renorm &10&112&94&0.05& \textbf{81} &0.05&\textbf{81} &1.34&\textbf{71}&2.65&69&\textbf{56}\\ \bottomrule 
\end{tabular}
\vspace{2em}
\caption{$\tgate$-count optimization evaluation. For PyZX and $\mathbf{PF}$ results, all other non-diagonal gate counts are left unchanged. Bolded entries in the $\mathbf{PF}$ columns denote entries which outperform PyZX, and the best entry (if it exists) in the +FastTODD columns is also bolded.}\label{tab:table}
\vspace*{-3em}
\end{table}

To test the impacts of our relational approach and the use of non-linear (classical) reasoning through the \textsc{Strengthen} algorithm to optimize $\tgate$-counts, we also performed circuit optimization experiments using a standard set of circuit benchmarks \cite{kv20}. We performed two sets of experiments --- ones to isolate the impact of \textsc{Strengthen} on phase folding, and ones to evaluate the overall effectiveness of our optimizations against other (non-monotone) circuit optimization algorithms. For the isolation experiments, we compared against PyZX \cite{pyzx} which implements a variant of phase folding in the ZX-calculus \cite{kv20}. We chose this version of the phase folding algorithm as it combines phase folding with Clifford normalization, and in particular reaches the theoretical limits of phase folding up to equalities and commutations of Clifford circuits \cite{optimal}. As Clifford computations implement \emph{affine} classical state transitions, our hypothesis is that our affine optimization algorithm $\mathbf{PF}_{\mathrm{Aff}}$ should match PyZX's optimizations in all cases. To test the overall efficacy of our optimizations in reducing $\tgate$-count, we also compared our optimizations against a selection of circuit optimizers (VOQC \cite{voqc}, PyZX \cite{pyzx}, FastTODD \cite{qpo}, QUESO \cite{queso}) which implement variants of phase folding (PyZX, VOQC), peephole-optimizations (VOQC, QUESO), and optimizations based on Reed-Muller decoding (FastTODD). As Reed-Muller optimizers apply an intrinsically different class of optimizations than phase folding \cite{tensor}, we also performed experiments comparing FastTODD after first either applying PyZX, or applying $\mathbf{PF}_{\textrm{Pol}}$. All circuits optimized by \textsc{Feynman} were validated for correctness using the method of \cite{a18}.

\Cref{tab:table} presents a summary of our experimental results, showing the isolation experiments and the overall best $\tgate$-count optimization results. As noted in \cite{queso}, peephole optimizations are generally ineffective at reducing $\tgate$-count as they are highly local. This is reflected in our results which show that either $\mathbf{PF}_{\textrm{Pol}}$, PyZX+FastTODD, or $\mathbf{PF}_{\textrm{Pol}}$+FastTODD always outperform the peephole optimizers, even when phase folding is also applied as a pre-processing step to QUESO. The peephole optimizers however often produce the best total gate counts, with FastTODD producing the worst overall gate counts due to a process of \emph{gadgetization} and re-synthesis. Overall, our results show that combining $\mathbf{PF}_{\textrm{Pol}}$ with Reed-Muller optimizations produces on average the best $\tgate$ counts, and usually outperforms PyZX+FastTODD when $\mathbf{PF}_{\textrm{Pol}}$ outperforms PyZX. We suspect some cases where $\mathbf{PF}_{\textrm{Pol}}$ did not improve on PyZX but PyZX+FastTODD outperformed $\mathbf{PF}_{\textrm{Pol}}$+FastTODD to be due to non-determinism in the FastTODD optimization.

Our isolation experiments confirm that \emph{affine} phase folding is able to match the $\tgate$ gate counts obtained by PyZX in every case. Thew few instances in which  $\mathbf{PF}_{\textrm{Aff}}$ outperforms PyZX are due to \textsc{Feynman}'s use of $\ket{0}$-initialized ancillas to further optimize $\tgate$-counts, which PyZX does not do. Our experiments further confirm that by adding in \emph{non-linear} equalities, phase folding is able to optimize away more $\tgate$ gates in many cases. We detail some of these cases below.

One such class of circuits where $\mathbf{PF}_{\mathrm{Pol}}$ consistently outperforms PyZX and $\mathbf{PF}_{\textrm{Aff}}$ are the $\Lambda_k(X)$  {\parfillskip0pt\par}

\setlength\intextsep{10pt}
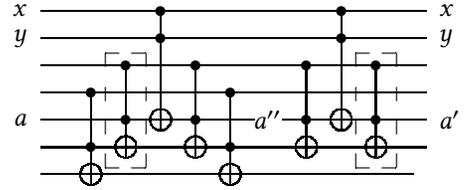
\begin{wrapfigure}{r}{0.45\textwidth}
\begin{center}
$
\Qcircuit @C=.5em @R=0.2em @!R {
		\lstick{x} & \qw & \qw & \qw & \qw & \ctrl{3} & \qw & \qw & \qw & \qw & \ctrl{3} & \qw & \qw & \qw & \rstick{x} \qw \\
		\lstick{y} & \qw & \qw & \qw & \qw & \ctrl{3} & \qw & \qw & \qw & \qw & \ctrl{3} & \qw & \qw & \qw & \rstick{y}\qw \\
		& \qw & \qw & \qw & \ctrl{2} & \qw & \ctrl{2} & \qw & \qw & \ctrl{2} & \qw & \ctrl{2} & \qw & \qw & \qw \\
		& \qw & \qw & \ctrl{2} & \qw & \qw & \qw & \ctrl{2} & \qw & \qw & \qw & \qw & \qw & \qw & \qw \\
		\lstick{a} & \qw & \qw & \qw & \ctrl{1} & \targ & \ctrl{1} & \qw & \push{a''}\qw & \ctrl{1} & \targ & \ctrl{1} & \qw & \qw & \rstick{a'}\qw \\
		 & \qw & \qw & \ctrl{1} & \targ & \qw & \targ & \ctrl{1} & \qw & \targ & \qw & \targ & \qw & \qw & \qw \\
		& \qw & \qw & \targ & \qw & \qw & \qw & \targ & \qw & \qw & \qw & \qw & \qw & \qw \gategroup{3}{5}{6}{5}{.7em}{--} \gategroup{3}{12}{6}{12}{.7em}{--}
	}
$
\end{center}
\caption{Implementation of the $4$-control Toffoli gate with $2$ dirty ancillas.
}\label{fig:barenco}
\end{wrapfigure}
\noindent 
benchmarks with \emph{dirty ancillas}, corresponding to an implementation due to \cite{bbcdmsssw95}. The $\Lambda_4(X)$ benchmark is reproduced in \cref{fig:barenco}. Observe that the equality $a' =a$ can be used to optimize away $4$ $\tgate$ gates from the indicated Toffolis.  Discovering the equality via generic means requires non-linear reasoning, as it relies on the intermediate equalities $a'' = a\oplus xy$ and $a'=a''\oplus xy$, and hence has previously been achieved by hand-optimization \cite{m16}. Our results match the $\tgate$-count scaling of $8(k-1)$ for the hand-optimized $\Lambda_k(X)$ implementation from \cite{m16} --- to the best of our knowledge, ours is the first circuit optimization to reach this $\tgate$-count via automated means.

Benchmarks where $\mathbf{PF}_{\mathrm{Pol}}$ outperformed both affine and \emph{quadratic} phase folding include Grover\_$5$ and FP-renorm. In the former case, $\mathbf{PF}_{\mathrm{Pol}}$ is able to reduce the $\tgate$-count to $0$, owing to the fact that rewriting the circuit's path integral gives the trivial transition $\ket{00\cdots 0}\mapsto \frac{1}{8}\sum_{\vec y\in \F_2^6} (-1)^{y_0}\ket{\vec y,000}$. It appears this is an error in the implementation of Grover's algorithm. The FP-renorm benchmark, taken from \cite{hht20}, was added specifically to provide a separation between quadratic and higher-degree optimizations. The circuit implements renormalization of the mantissa in floating point calculations, and it was shown that the effective $\tgate$-count can be reduced to $70$ by using a (programmer-supplied) state invariant which expresses a linear inequality over $\Z_k$. As expected, our results show a separation between $\tgate$-count optimizations possible with only quadratic equations ($81$) and those possible using higher-degree equations ($71$), which are generally needed when reasoning about integer arithmetic at the bit-level.

\section{Conclusion}

In this paper we have described a generalization of the quantum phase folding circuit optimization to quantum \emph{programs}. Our algorithm works by performing an relational analysis approximating the classical transitions of the program in order to determine where phase gates can be merged. We gave two concrete domains, one for affine relations based on \cite{elsar14}, and the other for polynomial relations based on Gr\"{o}bner basis methods. We further showed that abstract transformers for these domains can be generated by rewriting the circuit sum-over-paths. Our experimental results show that relational analysis is able to go improve upon existing circuit optimizations in many cases as well as discover some non-trivial optimizations of looping programs.

Many open questions and avenues for future work remain, particular generation of the \emph{most precise} transition relation for a quantum circuit. Other directions include expanding features of the analysis to include \emph{classical} logic such as branch conditions, and to experiment with the many existing classical techniques for relational analysis. A promising direction motivated by our experimental results is to explore techniques for \emph{degree-bounded} polynomial relations, for example \cite{ms04-2}. 
More broadly, this work raises the question of how else might \emph{classical} program analysis techniques --- particularly, numerical ones --- be applied in a quantum setting? One such avenue is to instead consider numerical invariants over the $\Z_N$, corresponding to the \emph{modular} basis $\{\ket{i} \mid i\in \Z_{2^n}\}$ of $n$ qubits which is frequently used in the analysis of quantum algorithms like Shor's seminal algorithm \cite{s94}. Yet another direction is to consider how the types of invariants we generate here may be applied to the \emph{formal verification} of quantum programs. 

\subsection{Related work}

\paragraph{Quantum circuit optimization}

The concept of \emph{phase folding} originated in the phase-polynomial optimization of \cite{amm14} and was later generalized from a re-synthesis algorithm to an in-place optimization in \cite{nrscm17}. The optimization was later re-formulated \cite{zc19,kv20} in the Pauli exponential and ZX-calculus frameworks, which extended the optimization to perform equivalent optimizations \emph{over all Pauli bases}. Both methods can be seen as phase folding ``up to Clifford equalities'', which is reflected in their experimental validation which curiously produced identical $T$-counts across all benchmarks.
Our method of rewriting the path integral, when restricted to \emph{affine} relations, can be seen as performing the same process of Clifford normalization, a fact which is mirrored in our experimental results for $\mathbf{PF}_{\textrm{Aff}}$ which coincide with \cite{zc19,kv20}. By using the symbolic nature of path integrals, we extend optimization further to \emph{non-Clifford} equivalences, notably those involving classical non-linearity.

If the problem of phase folding is relaxed to include \emph{re-synthesis}, another avenue of phase gate reduction opens up --- notably, techniques based on Reed-Muller codes \cite{am19}. Many recent phase gate optimization methods \cite{hc18,bbw20,tensor} make use of these techniques to often achieve lower $\tgate$-counts than those reported here. However, these methods rely on \emph{gadgetization} whereby $O(|C|)$ ancillas are added to the circuit $C$, and are generally orthogonal to phase folding \cite{kv20}. We do note that even recent gadget-based work \cite{tensor} was unable to find the optimizations of the dirty ancilla Toffoli gates found by our methods.

\paragraph{Quantum program optimization}

Unlike circuit optimization, quantum \emph{program} optimization is relatively unstudied. The work of \cite{thkh22} explicitly broaches the subject of dataflow optimization for quantum programs, developing a single static assignment (SSA) form for quantum programs and applying optimizations like gate cancellation on top of it. This notion differs from our notion of dataflow as their dataflow is explicitly classical, corresponding to the flow of \emph{quantum} values through a \emph{classical} control flow graph executing quantum operations at the nodes. In contrast, our model corresponds to \emph{classical} values with flow that is either \emph{either quantum (in superposition) or classical (out of superposition)}. In this sense, our work gives a notion of \emph{quantum dataflow} as \emph{dataflow in superposition}.

The work of \cite{hht20} shares many similarities with our approach. In particular, they use assertions expressed on the classical support of the quantum state to perform a number of local gate elimination optimizations. While their assertions are more expressive than the invariants we generate, as they include among other properties \emph{numerical inequalities} over the modular basis $\{\ket{i} \mid i \in \Z_{2^n}\}$, their assertions are user-supplied and not checked by a compiler. By contrast, our invariants are provably safe, generated automatically and as \emph{relational} invariants apply to non-local gate optimizations.

\paragraph{Relational program analysis}

Our work can be viewed as applying existing classical methods of relational, specifically algebraic, program analysis to the quantum domain. Our affine analysis directly uses the domain due to \cite{ks08} and \cite{elsar14}, which can be traced back to its origins in \cite{karr76}. Likewise, our polynomial analysis uses the theory of \emph{transition ideals} from the recent work \cite{ck24} and the algebraic foundations of \cite{rk04}. Further analogies with relational analysis can be made between our method of generating precise transition relations and the notion of computing the \emph{best symbolic transformer} in \cite{rsy04}, though the methods themselves share little similarities.

\paragraph{Quantum abstract interpretation}

Abstract interpretation has previously been applied to the problems of simulating circuits \cite{bpbv23} or verifying properties of circuits including separability \cite{p08} and projection-based assertions \cite{yp21,fl23}. The work of \cite{yp21} gave an abstract domain of products of subspaces on subsets of qubits and applied this to the verification of assertions in static (but large) quantum circuits. Later work \cite{fl23} defined a subspace domain $\subspace(\mathcal{H}_{\mathcal{Q}})$ consisting of subspaces of a Hilbert space $\mathcal{H}_{\mathcal{Q}}$ and gave operators for the interpretation of quantum WHILE programs, as well as translations to and from Hoare logics based on Birkhoff von-Neumann (subspace) logic. In contrast to these works, we abstract subspaces of the \emph{classical} states $\mathcal{C}_{\mathcal{Q}} = \F_2^{|\mathcal{Q}|}$, which crucially admits a polynomial-size basis and hence remains tractable. In this light, our methods can be seen as a further abstraction of the subspace domain, via the Galois connections $\powerset{\mathcal{H}_{\mathcal{Q}}} \galois{}{} \subspace(\mathcal{H}_{\mathcal{Q}})  \galois{}{} \subspace(\mathcal{C}_{\mathcal{Q}}).$
While the local subspace domain of \cite{yp21} likewise remains polynomial-size, our methods apply to general quantum programs and are able to prove the assertions for both the BV and GHZ test cases, as both circuits are precisely simulable using affine transformer semantics.

\section{Acknowledgments}
We wish to thank the anonymous reviewers for their comments and suggestions which have greatly improved the presentation of this paper. MA acknowledges support from Canada's NSERC and the Canada Research Chairs program.

% --------------------------------------------------------------------

\bibliographystyle{ACM-Reference-Format}
\bibliography{phasefold}

\appendix

\section{Additional experimental results}\label{sec:appendix}

In this appendix we give the full results of our experimental evaluation. \Cref{tab:tablecomplete} reports the total gate counts. Queso+PP refers to Queso with an additional phase folding pre-processing step via \textsc{Feynman}. Queso was allowed to run for $2$ hours and other tools were timed out after $2$ hours. The vast majority of optimization runs from all other tools completed within seconds. The FastTODD algorithm includes a phase folding step which is equivalent to the $\tgate$-count optimization in PyZX, as well as a gadgetization step which adds a large number (linear in the circuit length) of ancillas to the circuit. Only results computed by \textsc{Feynman} ($\mathbf{PF}_{\textrm{Aff}}$, $\mathbf{PF}_{\textrm{Quad}}$, and $\mathbf{PF}_{\textrm{Pol}}$) were formally verified for correctness.

\begin{landscape}
\begin{table}
\scriptsize
\begin{tabular}{lrrrrrrrrrrrrrrrrrrrr} \toprule
Benchmark & $n$ & \multicolumn{2}{c}{Original} & \multicolumn{2}{c}{voqc} & \multicolumn{2}{c}{queso} & \multicolumn{2}{c}{queso+PP} & \multicolumn{2}{c}{FastTODD} & \multicolumn{2}{c}{$\mathbf{PF}_{\mathrm{Aff}}$} & \multicolumn{2}{c}{$\mathbf{PF}_{\mathrm{Quad}}$}& \multicolumn{2}{c}{$\mathbf{PF}_{\mathrm{Pol}}$}& \multicolumn{2}{c}{$\mathbf{PF}_{\mathrm{Pol}}$+FastTODD} &\\ \midrule
 & &\# gates&\# $\tgate$&\# gates&\# $\tgate$&\# gates&\# $\tgate$&\# gates&\# $\tgate$&\# gates&\# $\tgate$&\# gates&\# $\tgate$&\# gates&\# $\tgate$&\# gates&\# $\tgate$&\# gates&\# $\tgate$&\\
\cmidrule(lr){3-4} \cmidrule(lr){5-6} \cmidrule(lr){7-8} \cmidrule(lr){9-10} \cmidrule(lr){11-12} \cmidrule(lr){13-14} \cmidrule(lr){15-16} \cmidrule(lr){17-18} \cmidrule(lr){19-20}&&&&&&&&&&&&&&&\\
Grover$\_5$ &9&831&336&626&172&748&290&660&176&7313&143&696& 148 &696&148 &586&\textbf{0}&10&\textbf{0}&\\
Mod $5\_4$ &5&63&28&54&16&45&10&33&10&54&\textbf{7}&52&8&52&8&52&8&61&\textbf{7}&\\
VBE-Adder$\_3$ &10&150&70&97&24&93&28&76&24&387&\textbf{19}&117&24&52&24&117&24&365&\textbf{19}&\\
CSLA-MUX$\_3$ &15&170&70&164&64&148&64&148&64&946&\textbf{39}&162& 60 &162& 60 &162&60&940&\textbf{39}&\\
CSUM-MUX$\_9$ &30&420&196&308&84&406&168&252&84&1526&\textbf{71}&336&84&336&84&336&84&3322&\textbf{71}&\\
QCLA-Com$\_7$ &24&443&203&332&99&335&135&272&95&1453&\textbf{59}&369& 94 &369& 94 &269&94&3161&61&\\
QCLA-Mod$\_7$ &26&884&413&753&259&784&321&644&245&10532&\textbf{159}&770&237&770&237&770&237&19339&161&\\
QCLA-Adder$\_10$ &36&521&238&449&162&466&198&394&164&2864&\textbf{109}&466&162&466&162&466&162&4626&\textbf{109}&\\
Adder$\_{8}$ &24&900&399&748&223&729&265&596&219&8569&\textbf{119}&737&173&737&173&737&173&9401&121&\\
RC-Adder$\_{6}$ &14&200&77&175&47&179&65&152&47&627&\textbf{37}&183&47&183&47&183&47&1185&\textbf{37}&\\
Mod-Red$\_{21}$ &11&278&119&227&73&229&75&196&73&1183&\textbf{51}&245&73&245&73&245&73&1853&\textbf{51}&\\
Mod-Mult$\_{55}$ &9&119&49&99&35&98&39&92&35&240&\textbf{17}&116&35&114&35&114&35&302&\textbf{17}&\\
Mod-Adder$\_{1024}$ &28&4285&1995&3419&1035&3905&1746&3197&1255&\multicolumn{2}{r}{TIMEOUT}&3363&1011&3337& 1005 &3115&\textbf{923}& \multicolumn{2}{r}{TIMEOUT}&\\
GF($2^4$)-Mult &12&225&112&192&68&198&82&176&68&667&\textbf{49}&194&68&194&68&194&68&632&\textbf{49}&\\
GF($2^5$)-Mult &15&347&175&292&115&299&135&273&115&1283&81&304&115&304&115&304&115&979&\textbf{75}&\\
GF($2^6$)-Mult &18&495&252&410&150&441&198&381&150&1607&113&414&150&414&150&414&150&2269&\textbf{111}&\\
GF($2^7$)-Mult &21&669&343&549&217&594&273&516&214&1745&155&553&217&553&217&553&217&2334&\textbf{141}&\\
GF($2^8$)-Mult &27&883&448&717&270&803&360&680&264&3861&205&709&264&709&264&709&264&3906&\textbf{203}&\\
GF($2^9$)-Mult &24&1095&567&887&351&943&459&835&351&4336&257&889&351&889&351&889&351&4021&\textbf{255}&\\
GF($2^{10}$)-Mult &27&1347&700&1084&410&1213&568&1008&410&6418&315&1088&410&1088&410&1088&410&5043&\textbf{313}&\\
GF($2^{16}$)-Mult &48&3435&1792&2715&1050&3295&1536&2613&1040&15641&\textbf{797}&2703&1040&2703&1040&2703&1040&19663&803&\\
GF($2^{32}$)-Mult &96&13562&7168&10594&4152&13050&6144&10563&\textbf{4128}&\multicolumn{2}{r}{TIMEOUT}&10562&\textbf{4128}&10562&\textbf{4128}&10562&\textbf{4128}&\multicolumn{2}{r}{TIMEOUT}&\\
Ham$\_{15}$ (low) &17&443&161&391&101&403&115&372&101&2679&\textbf{77}&391&97&391&97&391&97&2911&\textbf{77}&\\
Ham$\_{15}$ (med) &17&1272&574&946&266&1163&496&889&292&6315&\textbf{137}&901&212&901&212&895&210&8280&140&\\
Ham$\_{15}$ (high) &20&5308&2457&3832&1057&4768&2093&3988&1555&\multicolumn{2}{r}{TIMEOUT}&3969&1019&3914& \textbf{1007} &3927&\textbf{985}&\multicolumn{2}{r}{TIMEOUT}&\\
HWB$\_6$ &7&259&105&231&75&225&73&225&75&1328&\textbf{51}&237&75&237&75&237&75&1478&\textbf{51}&\\
QFT$\_4$ &5&179&69&158&67&176&67&168&67&308&\textbf{53}&173&67&172& 65 &172&65&927&54&\\
$\Lambda_3(X)$ &5&45&21&42&15&40&15&35&15&90&\textbf{13}&42&15&42&15&42&15&140&\textbf{13}&\\
$\Lambda_4(X)$ &7&75&35&69&23&65&23&55&23&218&\textbf{19}&69&23&69&23&69&23&256&\textbf{19}&\\
$\Lambda_5(X)$ &9&105&49&96&31&90&31&75&31&358&\textbf{25}&96&31&96&31&96&31&415&\textbf{25}&\\
$\Lambda_{10}(X)$ &19&255&119&231&71&215&71&175&71&1211&\textbf{55}&231&71&231&71&231&71&1558&\textbf{55}&\\
$\Lambda_3(X)$ (dirty) &5&60&28&52&16&48&16&38&16&112&\textbf{13}&52&16&52& 16 &52&16&110&\textbf{13}&\\
$\Lambda_4(X)$ (dirty) &7&114&56&98&28&89&28&68&28&246&23&99&28&97& 24 &97&24&292&\textbf{20}&\\
$\Lambda_5(X)$ (dirty) &9&170&84&144&40&130&60&98&40&442&33&146&40&142& 32 &142&32&528&\textbf{28}&\\
$\Lambda_{10}(X)$ (dirty) &19&450&224&374&100&335&160&248&100&1517&83&381&100&367& 72 &367&72&2361&\textbf{68}& \\
FP-renorm &10&266&112&254&94&264&108&251&94&2851&69&248& 81 &248& 81 &246&71&1425&\textbf{56}& \\ \midrule 
\end{tabular}
\vspace{2em}
\caption{Complete results of our experimental evaluation.}\label{tab:tablecomplete} 
\end{table}
\end{landscape}

\end{document}